\documentclass[12pt]{article}
\usepackage{amsmath,amssymb,bbold,epsfig}
\usepackage{amsfonts}
\usepackage{bbm}

\usepackage[colorlinks=true,linkcolor=red,citecolor=green,urlcolor=cyan,
bookmarks=true,bookmarksopen=true,pdfpagemode=None,pdfstartview=FitH]{hyperref}
\newcommand{\lambdabar}{
\setlength{\unitlength}{1pt}
\begin{picture}(6,7.5)
\put(-0.4,2.5){$-$}
\put(0,0){$\lambda$}
\setlength{\unitlength}{1cm}
\end{picture}}
\renewcommand{\vec}[1]{{\mathbf{#1}}}
\newcommand{\appsection}{\addtocounter{section}{1}\setcounter{equation}{0}
                         \renewcommand{\thesection}{\Alph{section}}
}
\renewcommand{\theequation}{\arabic{equation}}
\newcommand{\be}{\begin{equation}}
\newcommand{\ee}{\end{equation}}
\newcommand{\bea}{\begin{eqnarray}}
\newcommand{\eea}{\end{eqnarray}}
\begin{document}

\title{
\vglue -1.5cm
\rightline{\em\normalsize To the memory of A.B.~Migdal~~~~~}
\vskip 0.5cm
\Large \bf
Beta decay and other processes in strong electromagnetic fields}
\author{
{E. Kh. Akhmedov$^{a,b}$\thanks{email: \tt 
akhmedov@mpi-hd.mpg.de}~
} \\
{\normalsize\em $^a$Max--Planck--Institut f\"ur Kernphysik,
Postfach 103980} \\ {\normalsize\em D--69029 Heidelberg, Germany
\vspace*{0.15cm}}
\\
{\normalsize\em $^{b}$National Research Centre Kurchatov
\vspace*{-0.1cm}Institute}\\{\normalsize\em Moscow, Russia 
\vspace*{0.15cm}}
}
\maketitle 
\thispagestyle{empty} 
\vspace{-0.8cm} 
\begin{abstract} 
\noindent 
We consider effects of the fields of strong electromagnetic waves  
on various characteristics of quantum processes. After a qualitative 
discussion of the effects of external fields on the energy spectra 
and angular distributions of the final-state particles as well as on 
the total probabilities of the processes (such as decay rates and total 
cross sections), we present a simple method of calculating the 
total probabilities of processes with production of non-relativistic 
charged particles. Using nuclear $\beta$-decay as an example, we study 
the weak and strong field limits, as well as the field-induced $\beta$-decay 
of nuclei stable in the absence of the external fields, both in the tunneling 
and multi-photon regimes. We also consider the possibility of accelerating 
forbidden nuclear $\beta$-decays by lifting the forbiddeness due to the 
interaction of the parent or daughter nuclei with the field of a strong 
electromagnetic wave. It is shown that for currently attainable 
electromagnetic fields all effects on total $\beta$-decay rates are 
unobservably small.

\end{abstract}
%

\vspace{1.cm}
\vspace{.3cm}

\newpage

\section{\label{sec:intro}Introduction}
Study of quantum processes in intense electromagnetic fields is a very 
interesting subject. Strong external fields can help us to learn more 
about the properties of the involved particles and their interactions.   
Studying processes in strong fields may also have interesting implications for 
astrophysics and cosmology. Recently, there has been a renewed interest in this 
topic in connection with development of new powerful laser sources. 

In this paper I will discuss effects of strong external electromagnetic 
fields on various characteristics of quantum processes. 
In sec.~\ref{sec:qualit} a rather general qualitative analysis of 
these effects is given, whereas secs.~\ref{sec:allowed} and~\ref{sec:forbid}
are dedicated to a specific example -- nuclear $\beta$-decay in the field of 
a strong electromagnetic wave.

My interest in this topic was raised in the early 1980s by A.M. Dykhne, 
who called my attention to a paper published in the Physical Review Letters 
\cite{Beck1}. It was claimed in that paper that under the influence of 
electromagnetic fields of existing at 
that time powerful lasers $\beta$-decay of tritium can be significantly 
accelerated. Simple estimates I made did not confirm this conclusion, 
but at the same time I could not pinpoint a mistake in the calculation 
done in~\cite{Beck1}. The problem was that the calculation was 
very complicated and difficult to follow. It was based on the standard at 
that time procedure of infinite summation of partial probabilities 
corresponding to absorption from the external wave (or emission into it) of 
all possible numbers of photons. This motivated me to look for a simpler way 
of calculation of the total probabilities of quantum processes in the fields 
of intense electromagnetic waves. My quest was strongly supported by Victor 
Khodel, a colleague of mine at the Kurchatov Institute, who used to say 
that ``if there is a simple result, there must exist a simple way of 
obtaining it''. 

Eventually, I found a very simple way of calculating the total probabilities 
of quantum processes with emission of non-relativistic charged particles.
The method in particular allowed one to easily study all interesting limiting 
cases -- the weak and strong field limits as well as the case of the 
field-induced decay of a particle (or a nucleus) that is stable in the 
absence of the external field, both in the tunneling and multi-photon limits. 
The results were published in \cite{Akh1,Akh2}.  

While I was working on this subject, several papers appeared
\cite{tern1,Voloshin,nik1}, where the problem of $\beta$-decay in strong 
electromagnetic fields was re-investigated and it was shown that the results 
of \cite{Beck1} were erroneous (see also \cite{Beck2} and a later paper 
\cite{Beck3}). The analysis in 
\cite{tern1,nik1} was based on the same old summation technique, whereas the 
approach in Voloshin's 
paper \cite{Voloshin} was close in spirit (though not identical) to the one 
I was developing. 
In secs.~{\ref{sec:allowed} and~\ref{sec:forbid} I will 
discuss the method of \cite{Akh1,Akh2} using nuclear $\beta$-decay as an 
example, but it can actually be applied to a much wider class of problems.

Gratifyingly, in all considered limiting cases the results of direct 
calculations of the probability of $\beta$-decay in the field of a strong 
electromagnetic wave \cite{Akh1,Akh2} agreed perfectly well with my previously 
made estimates, often even including the numerical coefficients. I learned a 
great deal about how to analyze physics problems qualitatively and how to make 
simple estimates from A.B.~Migdal, both from my personal interactions with him 
and from his splendid book \cite{Migdal}. It is therefore a great pleasure and 
honour for me to write this article for the special issue of Yadernaya Fizika 
dedicated to the centennial anniversary of A.B.~Migdal's birthday.

\section{\label{sec:qualit}Qualitative considerations}
Consider quantum processes such as  
\begin{itemize}
\item[$\diamond$] ~~$A + B \to A + B $$ $ ~~~~~~~~~(scattering)

\item[$\diamond$] ~~$ A + B \to C + D + \dots $ ~(reactions)

\item[$\diamond$] ~~$ A \to B + C + \dots $ ~~~~~~~\,(decay)

\end{itemize}
How can external electromagnetic fields influence these processes? They can 
\begin{itemize}
\item 
Modify the differential characteristics of the process (i.e.\  
the energy spectra and angular distributions of final-state particles).

\item
Affect the total probabilities of the processes, such as total cross 
sections and decay rates. 

\item
Finally, new channels of reactions or decays, which were not available 
in the absence of the external fields, may open up.

\end{itemize}
Let us discuss qualitatively all these possibilities in turn. 

We will now make several assumptions that will be used throughout this paper. 
First, it turns out that the smaller the characteristic energies of the 
charged particles, the stronger the effects of external electromagnetic 
fields on the processes in which they are involved. For this reason we 
shall consider processes with 
non-relativistic charged particles.
We will assume that the system is subjected to the field of a monochromatic 
electromagnetic wave of frequency $\omega$ and electric field strength 
$\bf{E}$, and that the field is a 
low-frequency one: 
\be
\hbar \omega \ll \varepsilon_0\,, 
\label{eq:lowf}
\ee 
where $\varepsilon_0$ it the characteristic energy of the process.
The low-frequency limit very often also means that the wavelength of the 
photons of the external field $c/\omega$ is large compared to the 
characteristic length of the process $l_x$ (such as, e.g., the atomic 
size in photo-ionization processes or the nuclear radius in nuclear 
$\beta$-decay): $\omega l_x/c \ll 1$. The assumption that all the participating 
charged particles are non-relativistic allows us to concentrate only on 
effects of the electric component of the external field, whereas the 
condition $\omega l_x/c \ll 1$ implies that we can adopt the dipole 
approximation, in which the external electric field can be considered as a 
time-dependent uniform field ${\bf E}(t)$. Such a field can be 
conveniently described either in the Coulomb gauge
\be
A^\mu(t,{\bf x})=(0, {\bf A}(t))\,,\qquad {\bf E}(t) = -\frac{1}{c}
\frac{\partial {\bf A}(t)}{\partial t}\,
\label{eq:4vpot}
\ee
with the coordinate-independent vector-potential ${\bf A}(t)$, or in the 
scalar gauge
\be
A^\mu(t,{\bf x})=(\phi(t, {\bf x}), 0)\,,\qquad
\phi(t, {\bf x})=-{\bf E}(t){\bf x}\,,
\qquad {\bf E}(t) = -\nabla \phi \,.
\label{eq:4vpot1}
\ee
In different situations different gauges turn out to be most convenient 
for calculations; we will use the Coulomb gauge in sec.~\ref{sec:allowed} and 
the scalar gauge in sec.~\ref{sec:forbid} .

\subsection{\label{sec:diff}Differential characteristics} 
Strong external fields can modify energy spectra and angular distributions of 
particles produced in scattering, reaction or decay process. This happens due 
to the absorption by a charged particle of photons of the 
external field (or stimulated emission of photons into it).   

Free on-shell particles cannot absorb or emit photons due to 
energy-momentum conservation. However, a particle that undergoes a 
scattering which changes its momentum, 
or is produced in some process 
(such as an electron production in $\beta$-decay or emission of an 
electron from an atom due to photo-ionization or electron-impact 
ionization) can exchange photons with the external field. 
 
Let as a result of some process a non-relativistic particle of charge $e$ 
and mass $m$ be produced, and let its kinetic energy be 
$\varepsilon\le \varepsilon_0$, where $\varepsilon_0$ it the energy 
release in the process, i.e. the maximum kinetic energy available to the 
particle under consideration. The particle can receive some energy from the 
external field or give to the field a fraction of its energy. Let us estimate 
the corresponding energy $\Delta\varepsilon$. The exchange of the energy 
between the particle and the field effectively takes place during a 
characteristic time of order of the period of oscillations of the external 
field: $t_{char}\sim 1/\omega$ (the contribution of an integer number of full 
field periods $T=2\pi/\omega$ averages to zero). Therefore the momentum 
that the particle can obtain from the field is of order  
\be
\Delta k = e E_0 \,t_{char} \sim \frac {e E_0}{\omega}\,,
\label{eq:dk}
\ee
where $E_0$ is the amplitude of the electric field strength. 
For particles with a characteristic energy $\varepsilon_0$ (i.e. with the 
characteristic momentum $k_0 = \sqrt{2 m \varepsilon_0}$) we have   
\be
\frac{\Delta k}{k_0}\, =\, \frac {e E_0}{\sqrt{2 m \varepsilon_0}\,\omega}
\,\equiv\, \xi\,. 
\label{eq:xi}
\ee
This parameter characterizes, in particular, the modification of the angular 
distribution of the produced charged particle. If $\xi$ is not too large, 
during the characteristic time $t_{char}$ the particle moves over the distance 
$l\sim v_0 t_{char}=\sqrt{2\varepsilon_0/m}\,t_{char}$. The energy obtained 
by the particle from the external field is just the work of the field on the 
particle over the distance $l$, which gives   
\be
\frac{\Delta \varepsilon}{\varepsilon_0} ~\sim~ 
\frac{e E_0 l}{\varepsilon_0}
~=~\frac{e E_0}{\varepsilon_0} 
\Big(\sqrt{\frac{2\varepsilon_0}{m}}\,\frac{1}{\omega}\Big)
~=~ 2 \xi\,.
\label{eq:de}
\ee
This result is only valid assuming that $\xi\lesssim 1$; for $\xi \gg 1$ 
one has to take into account that the velocity of the particle increases with 
time and is no longer equal to its original velocity 
$v_0$. In this case 
\be
l\,\simeq\, v_0 t_{char} + \frac{eE_0}{m}\frac{t_{char}^2}{2}\,,
\label{eq:l2}
\ee
where $eE_0/m$ is the particle's acceleration in the external field. This 
yields 
\be
\frac{\Delta \varepsilon}{\varepsilon_0} ~\sim~ 
\frac{e E_0l}
{\varepsilon_0}
~=~\frac{e E_0}{\varepsilon_0}\Big(\sqrt{\frac{2\varepsilon_0}{m}}\,
\frac{1}{\omega}+\frac{e E_0}{m}\frac{1}{2\omega^2}\Big)
=2 \xi +\xi^2\,. 
\label{eq:de2}
\ee
Alternatively, this result can be obtained from $\Delta \varepsilon\simeq 
[(k_0+\Delta k)^2-k_0^2]/2m$ and eq.~(\ref{eq:xi}).

Thus, the modification of the differential characteristics of the process 
(energy spectra and angular distributions of the produced particles) is 
governed by  the parameter $\xi$ defined in eq.~(\ref{eq:xi}). For $\xi\gtrsim 
1$ the external field sizeably affects these quantities. 
The total number of photons absorbed from the field (or emitted into the 
field) for $\xi \lesssim 1$ can be estimated as 
\be
N_0 ~\simeq~\frac{\Delta\varepsilon}{\hbar \omega}~\simeq~\xi \cdot
\frac{2\varepsilon_0}{\hbar\omega}\,.
\label{eq:N0}
\ee
In the considered low-frequency limit (\ref{eq:lowf}) it can be very 
large even for not too strong fields, when $\xi$ is relatively small.  

The above estimates should be taken with some caution, though. The parameter 
$\xi$ diverges in the constant-field limit $\omega\to 0$; does this mean 
that $\Delta k$ and $\Delta \varepsilon$ will diverge as well? In fact, 
the above estimates of these quantities were made for the field of a plane 
electromagnetic wave, which in the limit $\omega\to 0$ goes into crossed 
uniform electric and magnetic fields of infinite space-time extension. For 
non-relativistic particles only the electric field matters, and in an 
electric field of infinitely large spatial size the quantity $\Delta 
\varepsilon$ can 
indeed formally become arbitrarily large. One should remember, however, that 
in reality all fields are limited in space and time and, in addition, the 
distance between the source and detector is finite. This leads to a natural 
cutoff in the expressions for $\Delta k/k_0$ and $\Delta \varepsilon/
\varepsilon_0$ in the limit $\omega\to \infty$: the characteristic time 
$t_{char}\sim 1/\omega$ should then be replaced by the smaller between 
the time scale of the field and the particle's time of flight between 
the source an the detector. The characteristic length $l$ in 
eq.~(\ref{eq:l2}) has to be modified accordingly.

Note that the parameter $\xi$ does not contain $\hbar$, i.e.\ is a purely 
classical quantity; therefore, the distortion of energy spectra and angular 
distributions of the final-state particles in external fields is a classical 
effect even when $N_0\sim 1$. In many interesting cases, however, and in 
particular for $\beta$-decay in strong laser fields, one has $\hbar\omega \ll 
\varepsilon_0$, so that for not too small $\xi$ one has $N_0\gg 1$. In those 
cases the exchange of energy between the system and the external field 
has a multi-photon nature. At the same time, the number of photons $N_1$ 
absorbed from the external field (or emitted into it) during the process 
of the formation of the emitted charged particle may be small. Indeed, 
the formation process is characterized by the time scale $t_0\sim 
\hbar/\varepsilon_0$; therefore the number of photons absorbed or emitted in 
the course of the particle production process is
\be
N_1\sim \frac{e E_0 \sqrt{2\varepsilon_0/m}\,t_0}{\hbar\omega}\,=\,
2\frac {e E_0}{\sqrt{2 m \varepsilon_0}\,\omega}\,=\, 2\xi\,. 
\label{eq:N1}
\ee
This means that $N_1\sim (\hbar \omega/\varepsilon_0) N_0 \ll N_0$, i.e.\ 
that the number of photons absorbed or emitted in the course of the 
production process is small compared to the total number of absorbed 
or emitted photons. The same is true for the energy change: the energy 
$\delta\varepsilon$ obtained by the particle from the wave (or given to 
the wave) during the process of its formation is small compared to the total 
change $\Delta \varepsilon$ of the particle's energy: 
\be
\delta\varepsilon\sim 
e E_0 \sqrt{2\varepsilon_0/m}\,t_0\,\simeq\,
\frac{\hbar \omega}{\varepsilon_0}\,
\Delta\varepsilon \ll 
\Delta \varepsilon\,. 
\label{eq:ineq1}
\ee
This means that for $\hbar\omega/\varepsilon_0\ll 1$ the distortions of 
the angular distributions and energy spectra of charged particles in 
external electromagnetic fields are mostly formed after their production 
process is already over. 

Thus, we have the following picture of how a quantum process in a 
low-frequency external electromagnetic field occurs. The whole process can be 
divided into two stages. In the first stage, charged particles are produced in 
a reaction, decay or scattering process; %
\footnote{Note that a scattering process $A+B\to A+B$ with some 
momentum transfer can be considered as destruction of particles in the 
initial state and their production in the final state.}
during the second stage, the produced particles exchange some energy with the 
external fields, and the observable energy spectra and angular distributions 
are formed. The two stages are to a large extent independent of each other, 
though the second stage is, of course, impossible without the first one.

{}From the above picture it follows that under certain conditions the 
differential cross sections of the processes should have a form of a product 
of two factors: the cross section of the process in the absence of the external 
field and the field-dependent factor describing the exchange of energy 
between the system and the external field. The second factor is practically 
independent of the character of the first stage of the process, i.e. is 
universal. 

The condition $\hbar\omega \ll\varepsilon_0$ which implies that the production 
or scattering of particles proceeds during times that are much shorter than the 
oscillation period of the external field, allows one to consider the 
first stage as a sudden perturbation, or ``jarring'' of the system in 
the presence of the external field. Such a concept was developed and 
analyzed in~\cite{DykhYud1}, where it was pointed out that  
for many different processes (such as e.g.\ stimulated bremsstrahlung, 
photo-ionization, inverse Compton scattering , etc.) in the field of an 
external electromagnetic wave the differential cross sections take the form
\be
\Big(\frac{d\sigma}{d\Omega}\Big)_n~\,\simeq~\left(\frac{d\sigma}{d\Omega}
\right)_0 J_n^2\left(\frac{e{\bf E}_0{\bf k}}{\hbar m\omega^2}\right)\,,
\label{eq:diff}
\ee
where $(d\sigma/d\Omega)_0$ is the field-free cross section, $n$ is the number 
of photons exchanged with the external field ($n>0$ corresponds to absorption 
of photons from the field and $n<0$ to their stimulated emission), and 
$J_n(z)$ is the Bessel function. From the well known relation 
$\sum_{n=-\infty}^\infty J_n^2(z)=1$ it then follows that the total cross 
section (i.e. the sum of the partial cross sections corresponding to all 
possible numbers of emitted or absorbed photons) coincides with the field-free 
one. In other words, in this approximation the total probability of the 
process is not modified by the external field, 

In fact, the condition $\hbar\omega \ll \varepsilon_0$ which ensures $N_1\ll 
N_0$ is a necessary but not sufficient condition for the field-independence of 
the first factor on the right hand side of eq.~(\ref{eq:diff}); for this, one 
would also have to require that the energy exchange between the system and the 
field in the process of the charged particle production,  
$\delta\varepsilon=N_1\hbar\omega$, be small compared to the characteristic 
energy of the process $\varepsilon_0$. We will discuss this condition in 
detail in the next subsection.

\subsection{\label{sec:total}
Field effects on total probabilities of the processes}

How can an external electromagnetic field affect the total probabilities 
(decay rates and cross sections) of quantum processes? There are 
essentially three possibilities: 
\begin{itemize}
\item[(i)] 
the fundamental interaction responsible for the process gets modified, 
leading to a modification of the transition operator; 

\item[(ii)] 
the matrix element of the process is altered due to a modification of 
the wave functions of the involved particles in the external field; 

\item[(ii)] the phase-space volume of the process gets changed.  

\end{itemize}
As an example of the first possibility, consider charged-current weak 
interaction processes (such as e.g.\ nuclear $\beta$-decay) in an external 
electromagnetic field.  
Virtual $W^\pm$ bosons which mediate these processes can interact with the 
field. However, each act of photon exchange between the $W$ boson and the 
field would lead to the appearance of an extra $W$-boson propagator in the 
amplitude of the process, and therefore would strongly suppress it due to the 
very large mass of the $W$ boson. A possible exception is the case when the 
field frequency is extremely high: $\hbar\omega\gtrsim m_W c^2$. However, the 
latter possibility would correspond to a process with the participation of a 
very hard $\gamma$-quantum, i.e. this would be a completely different process. 
Yet another possibility is when the external field, though a low frequency 
one, is very strong. It is easy to see, however, that in order for the field 
to produce a noticeable effect, the field strength has to be $E\gtrsim m_W^2 
c^3 /e\hbar\simeq 3.2\times 10^{26}$ V/cm,  
which is an extremely large value. It is not actually clear if such 
strong fields may exist in nature. 

The possibility (ii) can be realized, e.g., in the case of forbidden nuclear 
$\beta$-decay, where the interaction of nuclei 
with external electromagnetic fields 
may change the angular momentum of the wave functions of the 
initial and/or final nuclear states, 
thus lifting the forbiddeness of the $\beta$-transition. 
This possibility will be discussed in sec.~\ref{sec:forbid}. 

Let us now concentrate on the possibility (iii), i.e.\ on modification 
of the phase-space volume of the process. 

\subsubsection{Phase space change}

Assume that a charged particle produced in a quantum process obtains some 
energy from the external field. Can this actually increase the phase-space 
volume of the process and thus influence its total probability? 
At first sight, this seems to be impossible: Indeed, before the particle 
is produced, the field cannot affect it; after it has been produced, any 
change of its energy cannot affect the production probability. You can 
put the particle into a capacitor or accelerator and accelerate it to a 
very high speed -- this would not modify the production probability 
because the production process is already over.  

However, the above argument is purely classical, as it implies that the 
production process is instantaneous. Quantum mechanics tells us that in 
reality the production of a particle with an energy $\sim\varepsilon_0$ 
takes a finite time $t_0\sim \hbar/\varepsilon_0$. This is actually a 
formation time of the particle. The produced charged particle is 
not pointlike -- it is characterized by its de~Broglie wavelength:
\be
\lambdabar_D~\simeq~\frac{\hbar}{k_0}
~=~\frac{\hbar}{\sqrt{2 m \varepsilon_0}}\,.
\label{eq:DB1}
\ee
The formation time $t_0$ can be estimated as the time it takes for the 
particle's de~Broglie wave to emerge from the source (i.e.\ the time interval 
over which the particle moves over a distance of order of its de~Broglie 
wavelength):
\be
t_0 ~\sim~\lambdabar_D/v_0 ~=~\frac{\hbar}{\sqrt{2 m \varepsilon_0}}
\cdot \sqrt{\frac{m}{2 \varepsilon_0}}~=~\frac{\hbar}{2 \varepsilon_0}\,.
\label{eq:tau}
\ee

The energy obtained from the field {\it during the particle's 
formation time} can indeed increase the phase space volume of the 
process and affect its total probability (or cross section). The energy 
obtained after that  has no effect on the total probability -- it can 
only modify the energy spectra and angular distributions of the emitted 
particles. As discussed in sec.~\ref{sec:diff}, the latter effect is 
purely classical. At the same time, as we have just shown, modification 
of the total probabilities of processes in external electromagnetic fields is 
an inherently quantum effect.

Let us now estimate the energy obtained from the field by a particle during 
its formation time. It is given by the work done by the field on the particle 
over the characteristic distance $l_x$ equal to the particle's formation 
length, which should be of order $\lambdabar_D$. However, after the whole 
particle's de~Broglie wave has emerged from the source, the production process 
is already over, so for our estimate we take the characteristic length $l_x$ 
to be %
\be
l_x ~\simeq~\frac{\lambdabar_D}{2}\,.
\label{eq:lx}
\ee
The energy gain $\delta\varepsilon_D$ that affects the phase space volume of 
the process is then   
\be
\delta \varepsilon_D~\simeq~ e E_0 l_x~=~\frac{e E_0 \hbar}{2\sqrt{2 m 
\varepsilon_0}}\,,
\label{eq:deltaeps}
\ee
and the modification of the total probability is governed by the parameter 
\be
\chi~\equiv~\frac{\delta \varepsilon_D}{\varepsilon_0}~=~
\frac{e E_0 \hbar}{\sqrt{2 m \varepsilon_0} \,2 \varepsilon_0}\,.
\label{eq:chi}
\ee
Note that this parameter contains $\hbar$, as expected. The modification of 
total probabilities of quantum processes in the external fields would be 
substantial provided that $\chi\gtrsim 1$. 

Let us now give a slightly different argument for this. Assume 
that a charged particle is produced virtually with the energy ${\varepsilon_0
+\delta\varepsilon}$ instead of the energy ${\varepsilon_0}$ dictated by energy 
conservation. Such particle can only exist during a finite time interval  
${\tau_0\sim\hbar/ \delta\varepsilon}$; after that, it must be 
re-absorbed by its source. However, if during this time interval  
it receives the missing energy $\delta\varepsilon$ from the field, i.e.\ 
\be
e E \left(\sqrt{\frac{2\varepsilon_0}{m}}\,
\frac{\hbar}{\delta\varepsilon}\right)=\delta\varepsilon\,,
\label{eq:equal1}
\ee
it gets ``license to live'', i.e.\ can become real. To affect the 
production probability significantly, $\delta \varepsilon$ must be   
$\gtrsim \varepsilon_0$, which gives 
\be
\delta\varepsilon/\varepsilon_0~\sim~\chi ~\gtrsim~ 1\,, 
\label{eq:estim1}
\ee
i.e.\ the same condition as we found before. 

Let us stress that the parameter $\chi$ is independent of $\omega$ and 
therefore does not diverge in the constant-field limit $\omega\to 0$, 
unlike the parameter $\xi$ defined in~(\ref{eq:xi}). This is an  
important point. In ref.~\cite{Beck1} the tritium $\beta$-decay process 
\be
^3 {\rm H} ~\to~ ^3{\rm He} + e^- + \bar{\nu}_e
\label{eq:trit}
\ee
was considered
($\varepsilon_0=(M_{^3{\rm H}}-M_{^3{\rm He}}-m_e)c^2=18.6~\mbox{keV}$), 
and it was claimed that in the field of a plane electromagnetic wave the 
decay rate $W$ is 
\be
W \simeq W_0(\varepsilon_0)\cdot [1+c_0 \xi^2]\,,
\label{eq:WBeck}
\ee
where $W_0(\varepsilon_0)$ is the field-free decay rate and $c_0$ is a 
number of order unity. Were this result correct, it would mean that 
one could strongly influence tritium $\beta$-decay even by arbitrarily weak 
fields provided that their frequency is sufficiently small, which is 
obviously wrong.  Moreover, in the limit $\omega\to 0$ the result in 
eq.~(\ref{eq:WBeck}) clearly violates unitarity. 

We have pointed out in sec.~\ref{sec:diff} that the parameter $\xi$ 
governs the modification of the angular distributions and energy spectra 
of the particles in the external field, and that the divergence of $\xi$ in 
the limit $\omega\to 0$ does not pose any problems: the fact that all 
fields have finite space-time extensions introduces a natural cutoff for 
very small $\omega$. However, one can imagine a situation in which the 
field occupies a very large space-time region, and in general there is 
nothing wrong with the fact that, e.g., the energy gain of a charged 
particle in a constant electric field can be very large if the particle 
propagates very long distance in the field. For the total probability of the 
process, this argument would not work: the probability is always limited 
by unitarity, and therefore field-induced corrections to it cannot contain the 
field frequency $\omega$ in the denominator. From our analysis it follows that 
in relatively weak or moderate fields the tritium $\beta$-decay rate should be 
given by an expression similar to~(\ref{eq:WBeck}), but with $\xi^2$ replaced 
by $\chi^2$ (we will discuss this point in more detail in 
sec.~\ref{sec:estimtot}). Comparing eqs.~(\ref{eq:xi}) and~(\ref{eq:chi}) we 
find 
\be
\chi=(\hbar\omega/2\varepsilon_0)\xi\,,
\label{eq:chi2}
\ee 
i.e.\ for $\hbar\omega\ll \varepsilon_0$ one has $\chi \ll \xi$.   

Recall that the total number of emitted or absorbed photons $N_0$ 
and the number $N_1$ of photons exchanged with the field during the formation 
of the emitted charged particle are related to the parameter $\xi$ 
as $N_0\sim \xi (2\varepsilon_0/\hbar\omega)\gg \xi$ and $N_1\sim \xi$. 
At the same time, the corresponding total energy $\Delta \varepsilon$ obtained 
from the field or given to it and the energy $\delta\varepsilon_D$ exchanged 
with the field in the course of the particle's production can be expressed as 
\be
\Delta \varepsilon \simeq N_0\,\hbar\omega\simeq 2\xi \varepsilon_0\,,
\qquad \delta \varepsilon_D \simeq N_1 \hbar\omega\simeq \chi \varepsilon_0\,.
\label{eq:deltaeps2}
\ee

The above estimates of $\delta\varepsilon_D$ apply actually to the case of 
relatively weak external fields, when $\delta\varepsilon_D\lesssim 
\varepsilon_0$, i.e.\ $\chi\lesssim 1$. Let us now consider 

\subsubsection{\label{sec:strong}The strong field limit ($\chi \gg 1$)}
In deriving eqs.~(\ref{eq:DB1})-(\ref{eq:deltaeps}) we were assuming 
that the de~Broglie wavelength of the emitted charged particle is actually  
fixed by the energy release in the process $\varepsilon_0$, i.e.\ is 
field-independent. In the case of very strong external fields one has to 
take into account that the energy gain in the course of the particle 
formation can be large compared 
to $\varepsilon_0$, i.e. the de~Broglie wavelength of the produced particle 
is in general field dependent. Immediately after the particle production its 
distance from the source $l$ is small compared to its de~Broglie 
wavelength. As the particle moves away and its distance from the source 
increases, its de~Broglie wavelength decreases because the particle's momentum 
$k$ increases as $k\simeq k_0+e E_0 t$ (see fig.~\ref{fig:DB} 
where the particle is depicted by a small blob).
\begin{figure}[htb]
\hbox{\hfill
\hspace*{2.2cm}
{\includegraphics[width=4cm]{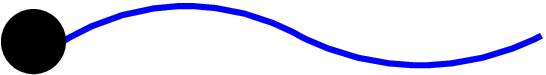}}
\hspace*{2.4cm}
{\includegraphics[width=4cm]{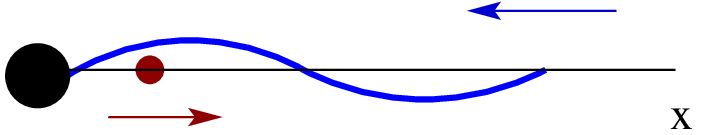}}
\hfill}
\caption{\label{fig:DB}Schematic representation of emission of the particle's 
de~Broglie wave from the source. Left: weak field case, the de~Broglie 
wavelength is constant. Right: strong field case, the particle gains 
significant energy and its the de~Broglie wavelength decreases as the particle 
moves away from the source.}
\end{figure}
The total probability of the process is only influenced by the energy 
that the particle gains over the distance of order of its de~Broglie 
wavelength, i.e.\ when the increasing $l$ and decreasing $\lambdabar_D$ 
``meet'' each other (more precisely, satisfy the condition 
$l=\lambdabar_D/2$, see~(\ref{eq:lx})). In the strong field limit we have   
\be
l ~\simeq~ v_0 t + \frac{eE_0}{m}\frac{t^2}{2}\,\simeq\, 
\frac{eE_0}{m}\frac{t^2}{2}\,,~~~
\label{eq:l3}
\ee
\be
\lambdabar_D \simeq \hbar/(k_0+e E_0 t)\,\simeq\, \hbar/(e E_0 t)\,, 
\label{eq:DB2}
\ee
and equating $l$ with $\lambdabar_D/2$ we obtain the ``meeting time'' 
$t_1$ and the ``meeting coordinate'' $l_1$: 
\be
t_1\simeq \left(\frac{m\hbar}{e^2 E_0^2}\right)^{1/3},\qquad
l_1= \frac{\hbar}{2e E_0 t_1}=\frac{\hbar}{2e E_0}\left(\frac{e^2 E_0^2}
{m\hbar}\right)^{1/3}. 
\label{eq:t0}
\ee
For the energy obtained by the particle from the field in the course of 
its production we find
\be
\delta\tilde{\varepsilon}_D\simeq e E_0 l_1 = 
\left(\frac{e^2 E_0^2 \hbar^2}{8m}\right)^{1/3}. 
\label{eq:deltaeps3}
\ee
The parameter that governs the modification of the total probability of 
the process in the strong field limit is therefore 
\be
\frac{\delta\tilde{\varepsilon}_D}{\varepsilon_0}
\simeq \frac{e E_0 l_1}{\varepsilon_0} = \left(\frac{e^2 E_0^2 
\hbar^2}{8m\varepsilon_0^3}\right)^{1/3}=\chi^{2/3}\,. 
\label{eq:deltaeps4}
\ee
Here the tilde over $\varepsilon_D$ is to distinguish this quantity 
in the strong field limit ($\chi\gg 1$) from the weak-field value given 
in eq.~(\ref{eq:chi}).

\subsection{\label{sec:estimtot}Estimates of total probabilities}
We have found that the parameters that describe the effects of external 
electromagnetic fields on the total probabilities of quantum processes 
are $\chi$ in the weak field limit and $\chi^{2/3}$ in the strong field case. 
Let us now try to quantify these effects. If the field-free rate of a 
process is $W_0(\varepsilon_0)$, the field-induced increase of the 
phase-space volume of the process would imply that the rate of the process 
in the presence of the field is given by 
$W\simeq W_0(\varepsilon_0+\delta\varepsilon_D)$. Let us first consider the 
weak field limit $\chi\ll 1$, i.e. $\delta\varepsilon_D\ll \varepsilon_0$. 
In this case we have 
\be
W~\simeq~
W_0(\varepsilon_0+\delta\varepsilon_D)~=~W_0(\varepsilon_0)
+W'_0(\varepsilon_0)\delta\varepsilon_D
+\frac{1}{2}W''_0(\varepsilon_0)(\delta\varepsilon_D)^2+\dots 
\label{eq:W1}
\ee
The first derivative term, which is linear in the field strength, vanishes 
upon the averaging over the phase of the field corresponding to the 
particle 
production time (or over the angle between ${\bf E}$ and the particle momentum 
${\bf k}$); the same is also true for all odd-order derivative terms 
in~(\ref{eq:W1}). Assuming that $W_0$ has a power-law dependence on 
$\varepsilon_0$ and taking into account that $\delta\varepsilon_D\simeq 
\chi\varepsilon_0$, we therefore obtain  
\be
W~\simeq~
W_0(\varepsilon_0)\big[1+\frac{1}{2}W''_0(\varepsilon_0)\chi^2+\frac{1}{4!}
W^{(IV)}_0(\varepsilon_0)\chi^4+\dots\big]. 
\label{eq:W2}
\ee
As an example, consider allowed nuclear $\beta$-decay with non-relativistic 
energy release, such as tritium $\beta$-decay. In this case 
$W_0(\varepsilon_0)\propto\varepsilon_0^{7/2}$ (see 
eq.~(\ref{eq:tritium}) below), and eq.~(\ref{eq:W2}) yields  
\be
W
~\simeq~
W_0(\varepsilon_0)\big[1+\frac{35}{8}\chi^2+\frac{35}{128}\chi^4+\dots\big].
\label{eq:W3}
\ee

This expression does not depend on the frequency of the external field, 
i.e.\ it actually corresponds to the limit $\omega\to 0$. How should the 
dependence on $\omega$ enter into the expression for $W$? The parameter 
governing this dependence is the ratio of the characteristic time scale of 
the process $t_x$ and the field period $T$, i.e.\ it is $\sim \omega t_x$. 
Due to the time reflection invariance of QED, the probability of the process 
can only depend on the even powers of this parameter. Thus, the coefficients 
of $\chi^{2n}$ in the expression for $W$ should actually be power series in 
$(\omega t_x)^2$. The coefficients in eq.~(\ref{eq:W3}) are just the leading 
(zero order) terms in these expansions. Note that in the considered case 
$\chi \ll 1$ we have $t_x\simeq t_0\equiv \hbar/2\varepsilon_0$, so that 
$(\omega t_x)^2=(\hbar\omega/2\varepsilon_0)^2$.

Consider now the strong field limit $\chi \gg 1$, in which 
$\delta\tilde{\varepsilon}_D\simeq \chi^{2/3}\varepsilon_0$. In this case 
to leading order one can neglect $\varepsilon_0$ in the expression  
$\varepsilon\simeq \varepsilon_0+\delta\tilde{\varepsilon}_D$ and 
expand $W_0(\varepsilon)$ in powers of 
$\varepsilon_0/\delta\tilde{\varepsilon}_D=\chi^{-2/3}$. This gives 
\be
W~\simeq~
W_0(\varepsilon_0+\delta\tilde{\varepsilon}_D)~\simeq 
W_0(\varepsilon_0\chi^{2/3})\big[1+{\cal O}(\chi^{-2/3})\big]\,.
\label{eq:W4b}
\ee
For allowed nuclear $\beta$-decay with emission of non-relativistic charged 
leptons, in which $W_0(\varepsilon_0)~\propto~\varepsilon_0^{7/2}$, we find  
\be
W~\simeq~
W_0(\varepsilon_0)\cdot \chi^{7/3}\cdot [1+{\cal O}(\chi^{-2/3})]\,.
\label{eq:W5}
\ee
Just like the weak-field expression~(\ref{eq:W3}), this result actually 
corresponds to the limit $\omega\to 0$. The $\omega$-dependence of the decay 
rate $W$ should be given by the power series in the parameter $(\omega t_x)^2$, 
where the characteristic time $t_x$ is now given by the ``meeting time'' $t_1$ 
defined in eq.~(\ref{eq:t0}), This yields 
\be
(\omega t_x)^2=\omega^2 \left(\frac{m\hbar}{e^2 E_0^2}\right)^{2/3}=
\chi^{-4/3}\Big(\frac{\hbar\omega}{2\varepsilon_0}\Big)^2.
\label{eq:tx2}
\ee
This means that the $\omega$ dependence of $W$ should only enter  
starting the term of order $\chi^{-4/3}$ in the expansion (\ref{eq:W5}), 
whereas the leading term and the term $\sim \chi^{-2/3}$ in the square 
brackets should be independent of the frequency of the external field.

Note that the leading term in (\ref{eq:W5}) ($\sim W_0(\varepsilon_0)
\chi^{7/3}$) could actually have been guessed. Indeed, since 
$W_0(\varepsilon_0)~\propto~\varepsilon_0^{7/2}$ and $\chi^{7/3}\propto 
\varepsilon_0^{-7/2}$, the leading term in (\ref{eq:W5}) is independent of 
$\varepsilon_0$. This is exactly as it must be in the strong field limit -- 
in this case the field-induced correction to the charged particle energy is 
large compared to the field-free energy, and the probability of the process 
should be approximately independent of the latter.

It is interesting to note that the fact that the leading-order term in $W$ is 
independent of $\varepsilon_0$ means that is is also independent of its  
sign. Therefore, eq.~(\ref{eq:W5}) holds true even in the case  
$\varepsilon_0<0$, which corresponds to the situation when in the absence of 
the external field the system is stable (or the reaction does not go 
because the available energy is below the threshold). 
For example, a sufficiently strong 
field can cause $\beta$-decay of an otherwise stable nucleus, such as $^3$He. 
The decay rate for this case is also described by eq.~(\ref{eq:W5}) 
provided that by $W_0$ and $\chi$ one understands $W_0(|\varepsilon_0|)$ and 
$\chi(|\varepsilon_0|)$.  
In other words, a very strong field ``pulls'' the electron (or positron) out  
of the nucleus irrespective of whether the nucleus was $\beta$-active or 
stable in the absence of the field.

\subsection{Weak-field limit for systems stable in the absence of 
the external field}
Consider now in more detail the case of negative energy release 
($\varepsilon_0<0$), e.g.\ field-induced $\beta$-decay of an otherwise stable 
nucleus. One can model such a situation 
by a particle bound in the potential well 
(fig.~\ref{fig:well}). In the absence of the external field the particle is 
in  a stationary bound state. The external dipole electric field adds 
the potential $U=-eE(t)x$ to the potential well, thus transforming it into a 
potential with a barrier. The state of the particle then becomes 
quasi-stationary; the particle can escape from the well either through 
the tunneling effect or due to a multi-photon ``ionization''.  
\begin{figure}[h]
\hbox{\hfill
\hspace*{1.4cm}
{\includegraphics[width=5.0cm]{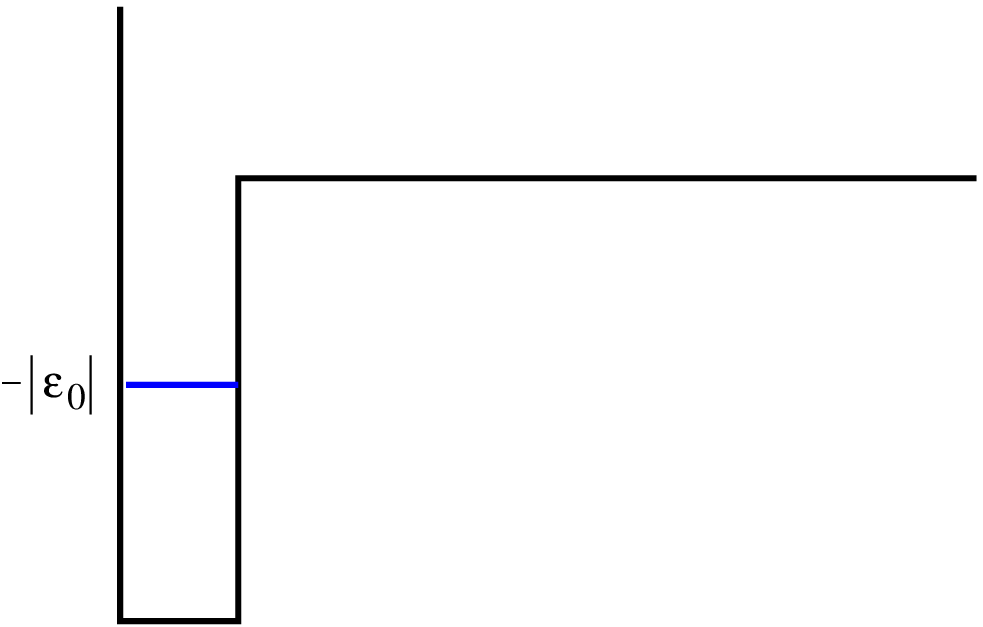}}
\hspace*{2.4cm}
{\includegraphics[width=5.0cm]{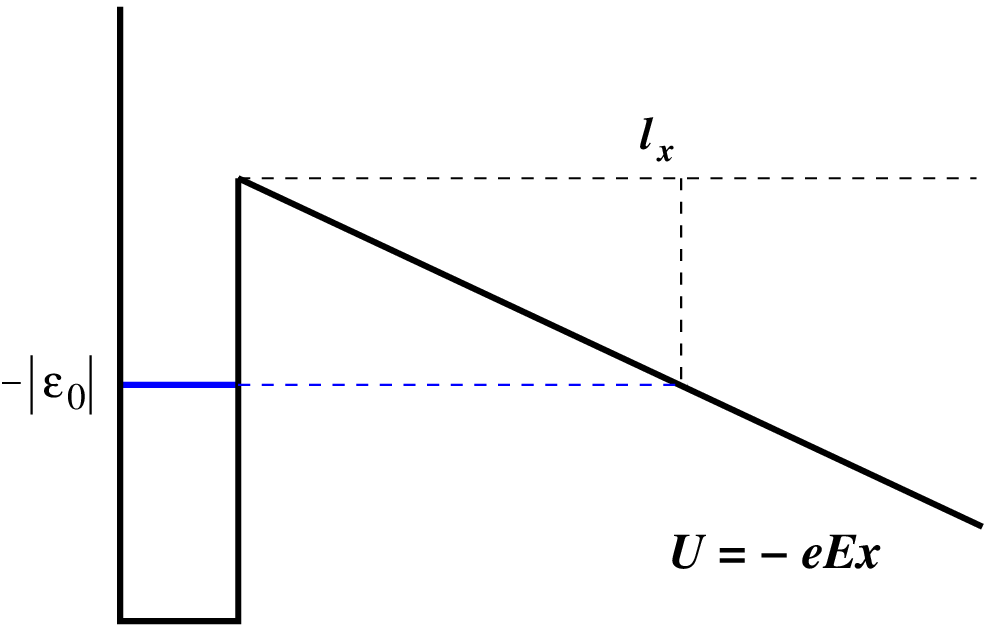}}
\hfill}
\caption{\label{fig:well} Left: a particle bound in a potential well; right: 
a particle in a well with a potential barrier. }
\end{figure}

Let us start with a qualitative analysis, similar to the one performed in the 
case $\varepsilon_0>0$. The $\beta$-particle can be emitted (along with the 
neutrino) virtually, and has to be re-absorbed after the time $t_0\sim\hbar/
|\varepsilon_0|$. On the other hand, the time it takes for the electron (or 
positron) to obtain the missing energy $|\varepsilon_0|$ from the external 
field is $t_2=\sqrt{2m |\varepsilon_0|}/e E$. If this time is smaller than or 
of the order of the ``time of virtual existence'' of the $\beta$ particle 
$t_0$, i.e.\ 
\be
t_2 = \sqrt{2m |\varepsilon_0|}/e E_0 ~\lesssim~\hbar/|\varepsilon_0|\,,
\label{eq:cond1}
\ee
the induced decay occurs with a large probability ($W/W_0(|\varepsilon_0|)
\gtrsim 1$). In the opposite case $t_2 \gg t_0$ the decay probability 
is 
strongly suppressed. 

The condition in eq.~(\ref{eq:cond1}) is equivalent to  $\chi(|\varepsilon_0|)
\gtrsim 1$ (see~(\ref{eq:chi})), i.e. corresponds to the strong field limit 
which was considered in the previous subsection. Let us now concentrate on the 
weak field limit $\chi(|\varepsilon_0|)\ll 1$. In this case the nature of the 
suppression of the rate of the induced $\beta$-decay will depend crucially on 
the value of $\omega t_2$. 
For $\omega t_2 \ll 1$ the external field is nearly static, and the 
field-induced $\beta$-decay proceeds as a tunneling effect. From 
fig.~\ref{fig:well} we find that the width of the barrier $l_2$ satisfies 
$e E_0l_2=|\varepsilon_0|$, i.e.\
\be 
l_2~=~|\varepsilon_0|/e E_0\,.
\ee
The parameter $t_2$ defined in (\ref{eq:cond1}) is then just the tunneling 
time. The condition $\chi \ll 1$ means that the WKB approximation can be used, 
and for the tunneling probability we find 
\be
W~\sim~\exp\Big[-\frac{2}{\hbar}\int_0^{l_2}\sqrt{2 m (|\varepsilon_0|-
e E_0 x)}\;dx\Big]~=~\exp\Big(-\frac{2}{3\chi}\Big).
\label{eq:WKB}
\ee
Note that this result is non-perturbative in the field strength $E_0$. 
Thus, in the low-frequency limit $\omega t_2 \ll 1$ the probability of 
the field-induced $\beta$-decay is exponentially suppressed. This result 
applies quite generally to all processes that do not occur in the absence of 
the external field. The pre-exponential factor in the expression for $W$ 
cannot be found from simple considerations like those presented above. 
The field frequency dependence of $W$ should, as usual, be given by the 
parameter $(\omega t_x)^2$, where $t_x$ now is the tunneling time $t_2$  
defined in eq.~(\ref{eq:cond1}), which gives 
\be
(\omega t_x)^2 = \xi^{-2}\,.
\label{eq:tx3}
\ee 

Consider now the weak field limit in the case $\omega t_2 \gg 1$. The 
potential barrier then oscillates fast on the time scale of the 
tunneling time, and the field-induced $\beta$-decay proceeds via 
the absorption of many photons by the $\beta$-particle, so that this 
particle goes over the barrier instead of tunneling through it. The 
minimum number of the absorbed photons is therefore $n_1=\lceil
|\varepsilon_0|/\hbar\omega\rceil$, i.e is the smallest integer that is 
$\ge |\varepsilon_0|/\hbar\omega$. The probability of the process should  
then depend on the field strength as 
\be
W \propto E_0^{2\lceil|\varepsilon_0|/\hbar\omega\rceil}\,,
\label{eq:n0}
\ee
i.e.\ it should exhibit a power-law rather than exponential suppression. 

\section{\label{sec:allowed}Allowed $\beta$ decay in the field of 
a strong wave}
We shall now go from simple estimates to direct calculations of the total 
probabilities of quantum processes in a field of a strong electromagnetic 
wave, taking allowed nuclear \mbox{$\beta$-decay} as an example. 

Let us first recall the decay rate calculation in the field-free case, 
According to the Fermi's Golden rule, the decay probability per unit time 
is given by 
\be
W_0(\varepsilon_0)~=~\frac{2\pi}{\hbar}\,|M_{fi}|^2 \int
\frac{d^3 k}{(2\pi\hbar)^3} \frac{d^3 p}{(2\pi\hbar)^3}\,\delta(M_i 
c^2-M_f c^2
-E_e-E_\nu)\,,
\label{eq:W6}
\ee
where ${\bf p}$ and ${\bf k}$ are the momenta of the emitted neutrino and 
electron, $E_e$ and $E_e$ are their full energies, and $M_i$ and $M_f$ are 
the masses of the parent and daughter nuclei. The energy release in the 
process is $\varepsilon_0=(M_i-M_f-m)c^2$, where $m$ is the electron mass 
(we neglect the recoil energy of the final-state nucleus). The matrix element 
of the process $M_{fi}=(G_F/\sqrt{2})\cos\theta_C {\cal M}_{fi}$, where $G_F$ 
is the Fermi constant, $\theta_C$ is the Cabibbo angle, and  ${\cal M}_{fi}$ 
is the nuclear matrix element (${\cal M}_{fi}={\cal O}(1)$ for 
allowed $\beta$-transitions). The $\delta$-function in~(\ref{eq:W6}) 
allows one to remove one of the energy integrations. Let us recall how 
it appears in eq.~(\ref{eq:W6}). The time dependent amplitude of the 
process is ${\cal A}(t)~\propto~e^{\frac{i}{\hbar}(E_f - E_i) t}$. The 
squared modulus of the integral of the amplitude over the production time 
therefore yields 
\be
\Big|\int_{-\infty}^{\infty}dt\, {\cal A}(t)\,\Big|^2
~\propto~\int_{-\infty}^{\infty}dt_1 \int_{-\infty}^{\infty}dt_2\, 
e^{\frac{i}{\hbar}(E_i - E_f) (t_1-t_2)}~=~T\!\cdot\! 2\pi \hbar\;
 \delta(E_i - E_f)\,,
\label{eq:ampsq}
\ee
where we have switched from the integration variables $t_1$ and $t_2$ to 
$(t_1+t_2)/2$ and $\tau\equiv t_1-t_2$. The integral over $(t_1+t_2)/2$ 
is trivial and yields the normalization time $T$ that has to be sent 
to infinity at the end of the calculations. The decay rate (i.e.\ the 
probability per unit time) is obtained by dividing the total probability 
by $T$ and therefore is $T$-independent. The integral over the difference of 
$t_1$ and $t_2$ gives the $\delta$-function $\delta(E_i-E_f)$, which helps 
perform the phase space integration in eq.~(\ref{eq:W6}). 

In the non-relativistic case ($\varepsilon_0\ll m c^2$) the 
direct calculation of~(\ref{eq:W6}) yields   
\be
W_0(\varepsilon_0)~=~\frac{|M_{fi}|^2}{\hbar^{7} c^{3}} 
(2m)^{3/2}(4/105\pi^3)\,\varepsilon_0^{7/2},
\label{eq:tritium}
\ee
i.e.\ $W_0\propto \varepsilon_0^{7/2}$, as was pointed out above.  

How can one calculate the decay rate in the presence of an external 
electromagnetic wave? Usually, the calculation is done by replacing the 
standard field-free wave function of the emitted electron with the exact 
solution of the Dirac equation in the field of a monochromatic 
electromagnetic wave -- the so called Volkov solution \cite{volk}. In the 
non-relativistic case and assuming that the dipole approximation is 
valid, one can use instead a much simpler wave function \cite{keldysh}, which 
in the Coulomb gauge (\ref{eq:4vpot}) it can be written as 
\be
\Psi_{\vec{k}}(\vec{r}, t)=\exp\left\{\frac{i}{\hbar}\vec{k} \vec{r}-
\frac{i}{2m\hbar}\int^t [\vec{k}-(e/c)\vec{A}(t')]^2\,d t' \right\}.
\label{eq:Keld1} 
\ee
This is an exact solution of the Schr\"odinger equation for an electron in 
a uniform electric field of arbitrary strength and arbitrary time 
dependence. Since in a non-stationary external field the $t$-dependence 
of the electron wave function is 
$\Psi_k(\vec{r}, t)~\propto~ e^{-i f(t)}~\ne~ e^{-\frac{i}{\hbar} E t}$, 
the calculation of the amplitude does not lead to an energy-conserving 
$\delta$-function. This simply reflects the fact that the energy of a system 
in a time-dependent external field is not conserved. 
For a periodic electric field $\vec{E}(t)$ with a period $T=2\pi/\omega$ 
the time-dependent factor $e^{-i f(t)}$ in the electron wave function 
can be expanded in a Fourier series: 
\be
e^{-i f(t)}=\sum_{n=-\infty}^{\infty} a_n e^{-i n \omega t}\,.
\label{eq:exp1}
\ee
Substitution of this into the expression for the transition amplitude 
would lead to a representation of the amplitude as a sum of partial amplitudes  
corresponding to the absorption from the field or stimulated emission into it 
of all possible numbers of photons. 

Consider a circularly polarized wave with the electric field strength 
\be
\vec{E}(t)~=~\{E_0 \sin\omega t, ~{}-E_0 \cos\omega t, ~0\,\}
\label{eq:field1}
\ee
By making use of the above Fourier expansion technique one can present the 
decay rate as 
\be
W=\frac{2\pi}{\hbar} |M_{fi}|^2\! \int\!\sum_{n=-n_0}^{\infty}\!
J_n^2\left(\frac{e E_0 k_\perp}{m \hbar \omega^2}\right)\delta(\varepsilon_0 
+n\hbar\omega-\varepsilon_K-\varepsilon-E_\nu)\,
\frac{d^3 k}{(2\pi\hbar)^3} \frac{d^3 p}{(2\pi\hbar)^3}\,,
\label{eq:W7}
\ee
where $k_\perp=\sqrt{k_x^2+k_y^2}$ is the component of the electron momentum 
in the plane of the field, $\varepsilon=k^2/2m$ is the kinetic energy 
of the electron, $\varepsilon_K=e^2E_0^2/2 m \omega^2$ is the mean kinetic 
energy of the oscillatory motion of the electron in the field of the wave, 
and $n_0=\lfloor(\varepsilon_0-\varepsilon_K)/\hbar\omega\rfloor$, i.e. 
the integer part of $(\varepsilon_0-\varepsilon_K)/\hbar\omega$. The 
appearance of the lower limit $\,-n_0$ in the sum in (\ref{eq:W7}) is 
related to the fact that the electron cannot emit into the field more energy 
than it has. In the limit $E_0\to 0$ only the term with $n=0$ in the sum 
survives and, since $J_0(0)=1$, the field-free probability 
(\ref{eq:W6}) is recovered.

The expression in eq.~(\ref{eq:W7}) was obtained by making use of the standard 
technique of Fourier expansion of the electron wave function. Now, let us look 
at this formula more closely. It contains an integral over the electron and 
neutrino 3-momenta of the sum of squared Bessel functions, whose arguments 
depend on the electron momentum and on the angle between this momentum and the 
direction of the external field strength. Moreover, because of the presence of 
the $\delta$-functions, the argument of the Bessel functions $J_n$ also 
implicitly depends on the index $n$ (through the dependence of $k$ on $n$). 
The summation extends from a finite value of $n$ to infinity. Obviously, it 
would be {\it very} difficult to calculate the probability of the process 
using this expression!

It is actually easy to understand why this calculational nightmare arises. 
The reason for this is a bad calculational approach. Indeed, each term 
in the sum in (\ref{eq:W7}) gives a partial probability of $\beta$-decay with 
absorption from the field (for $n>0$) or emission into it (for $n<0$) of $|n|$ 
photons. The energy conservation for each such processes is ensured by the 
corresponding $\delta$-function in the integrand. In fact, this means that the 
calculation is done in the energy representation. However, for a system in a 
non-stationary external field energy is not a good quantum number, therefore 
using the energy representation does not give us any advantage. Moreover, 
in order to calculate the total probability of the process, it is actually not 
necessary to calculate all partial probabilities and then sum them; the 
partial probabilities actually contain much more information than is needed. 
Equation (\ref{eq:W7}) therefore uses {\it excessive} information. 

The above arguments actually give us a hint of how an adequate calculation 
should be done. One should not make use of the Fourier expansion and  
calculate the partial probabilities. There will be no energy-conserving 
$\delta$-functions in this case, but the integrations over the 
electron and neutrino momenta can be performed without using 
$\delta$-functions. 
The idea of a simple calculation presented below is the following: 
\begin{itemize}
\item[1.]
In the calculation of the squared modulus of the integral of the amplitude 
of the process over the production time go from the variables $t_1$ and $t_2$ 
to  $(t_1+t_2)/2$ and $\tau=t_1-t_2$, as in the calculation that led to 
eq.~(\ref{eq:ampsq}).

\item[2.] Perform first the integration over $(t_1+t_2)/2$, then over the 
electron and neutrino \mbox{3-momenta}, and only at the very end do the 
integration over the difference of the times $\tau$.
\end{itemize}
Unlike in the field-free case, the integral over $(t_1+t_2)/2$ is 
not trivial now, but it still leads to the proportionality of the total 
probability of the process to the normalization time $T$. For the decay 
rate (i.e.\ probability per unit time) we find 
\[
W=\frac{1}{\hbar^2}\,|M_{fi}|^2 \int\frac{d^3 k}{(2\pi\hbar)^3} 
\int\frac{d^3 p}{(2\pi\hbar)^3}\hspace*{9.0cm}
\]
\be
~~~~~~~\times
\int_{-\infty}^{\infty}
J_0\Big(2\frac{e\,E_0\, k_\perp}{m \hbar\, 
\omega^2}\sin\frac{\omega\tau}{2}\Big)
\exp\Big\{\frac{i}{\hbar}\Big(\frac{k^2}{2m}+\frac{e^2E_0^2}{2 m \omega^2}
+E_\nu-\varepsilon_0\Big)\tau \Big\}\,d\tau\,.
\label{eq:W8}
\ee
Note that at this point it is still easy to go to the standard 
calculational technique. Indeed, by using the relation \cite{GrRyzh}
\be
J_0(2x\sin y)=\sum_{n=-\infty}^\infty J_n^2(x) e^{2iny}\,
\label{eq:J0}
\ee
one can reduce eq.~(\ref{eq:W8}) to~(\ref{eq:W7}).  

We now perform in~(\ref{eq:W8}) the momentum integrations,  
leaving the integration over $\tau=t_1-t_2$ 
to the end. 
The integration over the neutrino 3-momentum is trivial,%
\footnote{In this calculation the neutrino mass is neglected. We 
discuss the case $m_\nu\ne 0$ in sec.~\ref{sec:numass}.}
whereas the integral over the electron momentum can be readily 
done in cylindrical coordinates. The result is~\cite{Akh1} 
\be
W=\frac{\sqrt{\pi i}}{2^5\pi^4 \hbar^7 c^3}\,m^{3/2}\,|M_{fi}|^2\, 
(\hbar\omega)^{7/2}
\!\int_{-\infty}^{\infty}\frac{dx}{(x+i 0)^{9/2}}
\exp\Big\{-i\Big[\delta\!\cdot\!x+\gamma\Big(\frac{\sin^2 x}{x}-x
\Big)\Big]\Big\}\,.
\label{eq:W9}
\ee
Here the integration variable is $x=\omega (t_1-t_2)/2$, and we 
have introduced dimensionless parameters 
\be
\delta=\frac{2\varepsilon_0}{\hbar\omega}\,,\qquad
\gamma=\frac{e^2 E_0^2}{m \hbar\, \omega^3}\,.
\label{eq:deltagamma}
\ee
Note that in terms of $\gamma$ and $\delta$ the parameters $\xi$ and $\chi$ 
introduced in sec.~\ref{sec:qualit} are expressed as 
\be
\xi^2=\frac{\gamma}{\delta}\,,\qquad\quad \chi^2=\gamma/\delta^3\,.
\label{eq:xichi}
\ee 
Equation (\ref{eq:W9}) is the result we were looking for. Instead of 
the frightening expression (\ref{eq:W7}) we have now obtained a relatively 
simple single integral. It can be calculated numerically, but all the 
interesting limiting cases can actually be readily studied analytically.

Let us first note that our approximation (\ref{eq:lowf}) implies  
$\delta\gg 1$; therefore, the integral in (\ref{eq:W9}) gets its main 
contribution from the region $|x|\lesssim \delta^{-1} \ll 1$ and 
from the vicinities of the stationary phase points which lie outside this 
region. 
The stationary phase contributions are strongly suppressed 
fast-oscillating functions of the field strength $E_0$ which vanish upon 
averaging  over small fluctuations of this quantity; we will discuss 
their contribution later on. 
For small $|x|$ one can expand $\sin^2 x$ in the exponent in 
eq.~(\ref{eq:W9}) in powers of $x$. Keeping the first two terms in this 
expansion, we find   
\be
W=\frac{\sqrt{\pi i}}{2^5\pi^4 \hbar^7 c^3}\,m^{3/2}\,|M_{fi}|^2\, 
(\hbar\omega)^{7/2}
\!\int_{-\infty}^{\infty}\frac{dx}{(x+i 0)^{9/2}}
\exp\Big\{-i\Big[\delta\!\cdot\!x-\gamma\frac{x^3}{3}\Big]\Big\}.
\label{eq:W10}
\ee
The integral here is of Airy type. 

\subsection{\label{sec:weak}Weak-field limit (case $\varepsilon_0>0$)}

Consider first the limit 
\be
\gamma \ll \delta^3 \qquad\quad (\chi \ll 1)\,.
\label{eq:wfl}
\ee
For brevity, we shall call it `the weak-field limit' even though the 
field-dependent parameter $\gamma$ may actually be much greater than unity.  
For the values of $x$ 
satisfying $|x|\delta\lesssim 1$ 
one then has $\gamma |x|^3\ll 1$. Therefore in this regime one can expand 
\be
\exp\Big\{-i\Big[\delta\!\cdot\!x-\gamma\frac{x^3}{3}\Big]\Big\}
=e^{-i \delta x}\Big(1\,+\,i\gamma\frac{x^3}{3}\,+\,\dots\Big).
\label{eq:exp2}
\ee
Substituting this into (\ref{eq:W10}), we find 
\be
W~=~W_0(\varepsilon_0)\cdot\sum_{m=0}^{\infty}\frac{\Gamma(9/2)}
{m!\,3^m \,\Gamma(9/2-3m)}\chi^{2m}\,.
\label{eq:W11}
\ee
The first few terms in this expansion give 
\be
W~=~W_0(\varepsilon_0)\left[ 1+
\frac{35}{8}\chi^2\,+\,\frac{35}{128}\chi^4\,+\, \dots \right].
\label{eq:W12}
\ee
Comparing this with our estimates made in sec.~\ref{sec:qualit} (see 
eq.~(\ref{eq:W3})), we find a very good agreement. Not only 
the facts that the decay rate $W$ depends on the characteristics of the 
external field through the parameter $\chi$ and that for small $\chi$ 
it is a power series in $\chi^2$ were predicted correctly by our simple 
estimates -- even the numerical values of the coefficients of the $\chi^2$ and 
and $\chi^4$ terms were found correctly. This gives an additional 
{\it a posteriori} justification to our choice of the numerical coefficient in 
the expression for the characteristic length $l_x$ in eq.~(\ref{eq:lx}).  

The expression in eq.~(\ref{eq:W12}) actually correspond to the stationary 
field limit $\omega\to 0$. To obtain the dependence of $W$ on $\omega$ 
one has to retain next terms in the expansion of $\sin^2 x$  in the exponent 
in eq.~(\ref{eq:W9}). For the lowest order terms we obtain  
\be
W=W_0(\varepsilon_0)\left\{ 1+\frac{35}{8}\chi^2\left[1~-~\frac{1}{30}\!\left(
\frac{\hbar\omega}{2\varepsilon_0}\right)^2\right]\,+\,{\cal O}(\chi^4)
\right\}\,,
\label{eq:W13}
\ee
i.e.\ in the weak field limit the $\omega$-dependence enters into the 
expression for $W$ as an expansion in powers of $(\hbar\omega/2
\varepsilon_0)^2$, as it was anticipated in sec.~\ref{sec:qualit}. 

Let us now turn to the contributions to $W$ from the stationary phase points.
{}From (\ref{eq:W10}) one founds two stationary phase points:
\be
x_{1,2}=\pm(\delta/\gamma)^{1/2}=\pm \xi^{-1}\,.
\label{eq:stphase}
\ee
Their contribution to the decay rate is, to leading order,
\be
\delta W=W_0(\varepsilon_0)\frac{105}{16} \chi^4\cos(2/3\chi)\,.
\label{eq:statphase2}
\ee
Thus, the stationary phase contributions are suppressed at least as $\chi^4$ 
and are fast-oscillating functions of the field strength $E_0$ with zero 
average. 

\subsection{\label{sec:strong1}Strong field limit ($\chi \gg 1$)}

In the strong field limit $\chi \gg 1$ (which corresponds to $\gamma \gg 
\delta^3$) one can, instead of~(\ref{eq:exp2}), use the expansion  
\be
\exp\left\{-i\left[\delta\!\cdot\!x-\gamma\frac{x^3}{3}\right]\right\}
=e^{i \gamma\frac{x^3}{3}}\left(1\,-\,i\delta 
x\,-\frac{\delta^2}{2}+\,\dots\right).
\label{eq:exp3}
\ee
Substituting this into eq.~(\ref{eq:W10}) yields 
\be
W~=~W_0(\varepsilon_0)\,\chi^{7/3}\cdot
\frac{35}{144}
\frac{3^{5/6}}{\sqrt{\pi}}
\sum_{m=0}^{\infty}\frac{(-3)^m\,\chi^{-2m/3}}{m!}\sin\left(\frac{2\pi 
m}{3}+\frac{\pi}{6}\right)\Gamma\left(\frac{m}{3}-\frac{7}{6}\right). 
\label{eq:W14}
\ee
This result holds for $\beta$-active nuclei ($\varepsilon_0>0$). For 
nuclei that are stable in the absence of the field ($\varepsilon_0 < 0$) 
the decay rate can be obtained from eq.~(\ref{eq:W14}) by replacing $(-3)^m\to 
3^m$, $W_0(\varepsilon_0)\to W_0(|\varepsilon_0|)$ and $\chi(\varepsilon_0)\to 
\chi(|\varepsilon_0|)$.  

To obtain the dependence of the decay rate on the field frequency $\omega$ one 
has, as usual, to retain the next terms in the expansion of $\sin^2 x$ in 
powers of $x$ in eq.~(\ref{eq:W9}). This yields   
\[
W=W_0(|\varepsilon_0|)\,\chi^{7/3}
\cdot\left\{
\frac{5\!\cdot\! 3^{5/6}\!\cdot\!\Gamma(5/6)}{8\sqrt{\pi}}\,\pm\, 
\frac{7\!\cdot\! 3^{1/6}\!\cdot\!\Gamma(1/6)}{16\sqrt{\pi}}\chi^{-2/3} 
\right.
\]
\be
\left.
+\,\frac{35\sqrt{3}}{48}\chi^{-4/3}\left[1\,+\,\frac{2}{15}
\left(\frac{\hbar\omega}{2\varepsilon_0}\right)^2\right]\,+\,{\cal 
O}(\chi^{-2})\right\}.
\label{eq:W15}
\ee
Here the upper and lower signs correspond to $\varepsilon_0>0$ and 
$\varepsilon_0<0$, respectively. 

Let us compare this result with the estimates made in sec.~\ref{sec:qualit}. 
It was predicted there that in the strong field limit 
$W\simeq W_0(\varepsilon_0)\chi^{7/3}$ times a power series in $\chi^{-2/3}$ 
(see~(\ref{eq:W5})). Equation~(\ref{eq:W15}) confirms this result, 
except that 
the leading term has an extra constant factor $C_1$. However, for 
the numerical value of this factor we have 
\be
C_1=\frac{5\!\cdot\! 3^{5/6}\!\cdot\!\Gamma(5/6)}{8\sqrt{\pi}}=0.9943\,,
\label{eq:C1}
\ee
which is very close to 1. Thus, the predictions based on simple estimates 
are confirmed once again. The first two terms in the curly brackets 
in eq.~(\ref{eq:W15}) are $\omega$-independent, and the dependence of $W$ 
on $\omega$ comes through the positive powers of the quantity 
$\chi^{-4/3}(\hbar\omega/2\varepsilon_0)^2$, 
again in full agreement with the results of our qualitative analysis in 
sec.~\ref{sec:qualit}. As was discussed there in detail, the strangely-looking 
dependence of $W$ on fractional powers of $\chi$ is a consequence of the fact 
that in strong fields the electron's de~Brogle wavelength is field-dependent.

\subsection{\label{sec:weak2}The case of weak fields and 
$\varepsilon_0<0$ }
Consider now the case of relatively weak external fields and nuclei stable 
in the absence of the field ($\varepsilon_0<0$). As was discussed in  
sec.~\ref{sec:qualit}, there are two essentially different regimes of 
the induced $\beta$-decay in this case, the tunneling regime and the 
multi-photon one, depending on whether the field frequency $\omega$ is 
small or large compared to the inverse tunneling time $t_2^{-1}=
e E_0/\sqrt{2m|\varepsilon_0|}$. Both cases can be obtained from the same 
master equation (\ref{eq:W9}). In the low-frequency limit $\omega 
t_2=\xi^{-1}\ll 1$ in the zeroth order in this parameter one can 
actually make use of a simpler formula (\ref{eq:W10}). The calculation of 
the integral in (\ref{eq:W10}) in the steepest descent approximation yields  
\be
W(\chi)=W_0(|\varepsilon_0|)\frac{105}{32\sqrt{\pi}}\,\chi^4 \,e^{-2/3\chi}
\sum_{k=0}^{\infty}(-\chi)^{k} \sum_{n=0}^{2k}\frac{\Gamma(9/2+2k-n)
\Gamma(k+n+1/2)}{n!\, (2k-n)!\, \Gamma(9/2)}\,. 
\label{eq:W16}
\ee
To take into account the $\omega$ dependence of the decay rate, one has 
to retain the higher-order terms in the expansion of $\sin^2 x$ in the 
exponent in eq.~(\ref{eq:W9}). This gives, to leading order, 
\be
W(\chi)=W_0(|\varepsilon_0|)\frac{105}{32}\left(1+\frac{5}{9\xi^2}\right)\,
\chi^4 \,\exp\left[-\frac{2}{3\chi}\left(1-\frac{1}{15\xi^2}\right)\right].
\label{eq:W17}
\ee
 
In the multi-photon regime ($\omega t_2=\xi^{-1}\gg 1$), one can find the 
probability of the process by calculating the integral in (\ref{eq:W9}) in 
the saddle point approximation. The simpler expression (\ref{eq:W10}) cannot 
be used in this case since a region of complex $x$ with large modulus 
is important in the integral. The saddle point is found from the 
transcendental equation 
\be
\sinh^2 z - (\cosh z - z^{-1}\sinh z)^2=\xi^{-2}\,,
\label{eq:saddle}
\ee
where $z=ix$. 
Solving it approximately in the limit $\xi \ll 1$, we find 
\be
W~\simeq~W_0(|\varepsilon_0|)\frac{105}{32}\left(\frac{\hbar\omega}
{2|\varepsilon_0|}\right)^4 \left(\ln\frac{2m|\varepsilon_0| \omega}
{eE}\right)^{-9/2}\left(\frac{eE}{2m|\varepsilon_0|\omega}
\right)^{\frac{2|\varepsilon_0|}{\hbar\omega}-1/2}. 
\label{eq:W18}
\ee

Thus, we see that, in accordance with the qualitative analysis of 
sec.~\ref{sec:qualit}, in the weak-field limit $\chi \ll 1$ the probability 
of the field-induced $\beta$ decay is exponentially suppressed in the 
low-frequency case and exhibits a power-law suppression in the 
high-frequency limit.    

\subsection{\label{sec:numass}Neutrino mass effects}
Up to now, in our calculations we have been neglecting the neutrino mass 
$m_\nu$. In the absence of external fields, the only characteristic 
of $\beta$-decay with the dimension of energy is the energy release 
$\varepsilon_0$, and it is always much greater than 
$m_\nu c^2$. This justifies the neglect of $m_\nu$ in calculations of 
the total $\beta$-decay rates in the absence of external fields.

In the case of $\beta$-decay in a field of monochromatic wave, one can 
construct other quantities with the dimension of energy, which depend on the 
field strength and frequency and which have to be compared to $m_\nu c^2$. 
If any characteristics of the process should turn out to depend on the ratios 
of $m_\nu c^2$ and these field-dependent parameters, study of $\beta$-decay 
in electromagnetic fields could provide important information on the value 
of the neutrino mass. %
\footnote{Note that the effective neutrino mass $m_\nu$ that can be probed in 
$\beta$-decay is expressed through the masses of neutrino mass eigenstates 
$m_i$ and the elements of the leptonic mixing matrix $U$ as $m_\nu=\sum_i 
|U_{ei}|^2 m_i$.} 
Using $E_\nu=(p^2+m_\nu^2)^{1/2}$ in eq.~(\ref{eq:W8}) and following the same 
steps as those that led to (\ref{eq:W9}), we obtain~\cite{Akh1}
\[
W=-\frac{\sqrt{\pi i}}{2^4\pi^4 \hbar^7 c^3}\,m^{3/2}\,|M_{fi}|^2\, 
(m_\nu c^2)^2 (\hbar\omega)^{3/2}
\qquad\qquad\qquad\qquad\qquad\qquad\qquad\qquad
\]
\be
\times\int_{-\infty}^{\infty}\frac{dx}{(x+i 0)^{5/2}}
K_2\left(-2i\frac{m_\nu c^2}{\hbar\omega}x\right)
\exp\Big\{-i\Big[\delta\!\cdot\!x+\gamma\Big(\frac{\sin^2 x}{x}-x
\Big)\Big]\Big\}\,.
\label{eq:W19}
\ee
This equation can be studied by the same methods as those that were applied to 
eqs.~(\ref{eq:W9}) and~(\ref{eq:W10}). The analysis \cite{Akh2} shows 
that the dependence of the $\beta$-decay rate $W$ on the neutrino mass has the 
same nature as its dependence on the field frequency $\omega$, i.e.\ it enters 
through the parameter $(m_\nu c^2 t_x/\hbar)^2$, where $t_x$ is the 
characteristic time scale of the process. As was discussed above, $t_x$ is 
different in different regimes. The parameter $(m_\nu c^2 t_x/\hbar)^2$ turns 
out to be very small in all cases except for $\varepsilon_0<0$, $\chi \ll 
1$, when $t_x$ is the tunneling time $t_2$ defined in~(\ref{eq:cond1}). In 
this case the corrections to the probabilities of field-induced $\beta$-decay 
due to non-vanishing neutrino mass can be sizable. However, the decay 
probabilities themselves  are extremely small then, making the neutrino mass 
effects unobservable. Thus, unfortunately $\beta$-decay in external 
electromagnetic fields cannot tell us much about the neutrino mass, at least 
as far as the total probabilities are concerned. Note that interesting effects 
on the spectra of $\beta$-electrons may still be possible. 

\section{\label{sec:forbid}Forbidden $\beta$ decay in strong fields -- 
lifting the forbiddenes}
As was pointed out in sec.~\ref{sec:total}, one possible mechanism of 
modification of the total probabilities of quantum processes in  
electromagnetic fields is modification of their transition matrix elements. 
We shall discuss now a particular example of this -- acceleration of 
forbidden nuclear $\beta$-decay due to the interaction of the parent or 
daughter nuclei with the field.  

When considering allowed $\beta$ decay in external fields we were taking 
into account only the interaction of the produced electron or 
positron with the field; the interaction of the involved nuclei was 
ignored because they are very heavy. In the case of forbidden 
$\beta$-decay, however, there exists an enhancement mechanism which 
requires taking the interaction of nuclei with the field into account. 
If the parent or daughter nucleus absorbs from the field (or emits into 
it) one or more photons in the course of $\beta$-decay, the angular momentum 
of the initial or final nuclear state may change, and the forbiddeness 
may be lifted. 

Let us discuss this mechanism in more detail. Let a nucleus undergo a 
forbidden \mbox{$\beta$-decay} from its ground state $|i\rangle $ to the 
ground state 
$|f\rangle$ of the daughter nucleus. Assume that the parent nucleus has an 
excited state $|1 \rangle$ whose quantum numbers permit an allowed 
\mbox{$\beta$-transition} $|1\rangle \to |f\rangle$. The parent nucleus can 
then undergo a virtual transition to the state $|1 \rangle$ by absorbing 
one or more photons from the external field (or by emitting them into 
the field), followed by the allowed $\beta$-transition. An analogous 
situation will obtain if the daughter nucleus has an excited state 
$|2 \rangle$ with quantum numbers permitting an allowed $\beta$-decay 
$|i\rangle \to |2\rangle$, followed by the electromagnetic $|2\rangle \to 
|f\rangle$ transition under the influence of the external field. %
\footnote{We are assuming here that the excitation energy of the state 
$|2\rangle$ is larger than the energy release $\varepsilon_0$ of the ground 
state $\to$ ground state $\beta$-transition. Otherwise the allowed 
$\beta$-decay $|i\rangle \to |2\rangle$ followed by a cascade of 
$\gamma$-transitions $|2\rangle \to |f\rangle$ would be possible even in 
the absence of the external field.}

Let us now concentrate on the simple case of unique first-forbidden 
$\beta$-transitions for which the selection rules are $\Delta J^{\Delta \pi}=
2^{-}$. Higher-order forbidden transitions can be examined similarly. 

For unique first-forbidden transitions, the state $|1 \rangle$ must be 
related to the ground state $|i \rangle$ of the parent nucleus by the  
electric dipole ($E1$) transition. The $|1 \rangle \to |f \rangle$ 
transition will then be the allowed Gamow-Teller transition with 
$\Delta J^{\Delta \pi}=1^{+}$. An analogous situation would arise for 
transitions through the virtual state $|2 \rangle$ of the daughter 
nucleus. It is readily seen that any other possibility (e.g., 
electromagnetic $M2$ transition and allowed Fermi transition with 
$\Delta J^{\Delta \pi}=0^{+}$) would result in a much smaller matrix element. 

Let us estimate the matrix element $M_{ind}$ of the induced process 
corresponding to the relaxation of the forbiddeness in external fields. 
We shall denote the energy differences between the states $|1 \rangle$,
$|i \rangle$ and $|2 \rangle$, $|f \rangle$ as
\be
\Delta \varepsilon_1\equiv \varepsilon_1-\varepsilon_i\,,\qquad
\Delta \varepsilon_2\equiv \varepsilon_2-\varepsilon_f\,.
\label{eq:deltas}
\ee
The admixture of the state $|1 \rangle$ to the ground state $|i \rangle$ 
in the external field of frequency $\omega \ll \Delta \varepsilon_1$ is 
characterized by the parameter $e E_0 d_{1i}/\Delta\varepsilon_1$, where 
$d_{1i}$ is the dipole matrix element of the corresponding electromagnetic 
transition. Similarly, the admixture of the state $|2 \rangle$ to the ground 
state $|f \rangle$ of the daughter nucleus  
is characterized by the parameter $e E_0 d_{2f}/\Delta\varepsilon_2$.
To estimate the matrix element $M_{ind}$, we replace the dipole matrix 
elements $d_{1i}$ and $d_{2f}$ by the nuclear radius $R$ and obtain
\be
M_{ind}\sim \frac{e E_0 R}{\Delta\varepsilon_{1,2}}\cdot1\,,
\label{eq:Mind}
\ee
where for the estimate we have assumed that the matrix elements of allowed 
$\beta$-transitions are of order unity (in reality, for intermediate and heavy 
nuclei they are somewhat smaller). 

Note that the ``forbiddeness lifting 
parameter'' $e E_0 R/\Delta \varepsilon_{1,2}$ does not depend on the 
field frequency $\omega$ in the limit $\omega \ll \Delta \varepsilon_{1,2}$. 
This is in contrast with the results of refs.~\cite{Reiss1,Reiss2}, where 
it was claimed that the forbiddeness can be relaxed if the parameter 
$z^{1/2}\equiv e E_0 R/\omega$ becomes sizeable. As was 
discussed in sec.~\ref{sec:qualit}, the dependence of the total probabilities 
on the parameters that diverge in the limit $\omega\to 0$ is not 
admissible from the physical point of view; it was shown in 
\cite{Akh2} that the results of refs.~\cite{Reiss1,Reiss2} were a 
consequence of some unjustified approximations adopted in those papers, 
which, in particular, led to a breakdown of gauge invariance. As a 
result, the probabilities of forbidden nuclear $\beta$-decay in external 
electromagnetic fields were overestimated in~\cite{Reiss1,Reiss2} by many 
orders of magnitude. 

The dependence of the induced matrix element $M_{ind}$ on the external field 
frequency can become important for large enough $\omega$, when $\omega\sim 
\Delta\varepsilon_{1,2}$; in this case in eq.~(\ref{eq:Mind}) $\Delta
\varepsilon_{1,2}$ in the denominator has to be replaced by $\Delta
\varepsilon_{1,2}\pm \omega$. In particular, for $\omega=\Delta
\varepsilon_{1,2}$ a resonant enhancement of the process becomes possible; 
the denominator of $M_{ind}$ in eq.~(\ref{eq:Mind}) would then just contain 
the width of the corresponding excited state, $\Gamma_1$ or $\Gamma_2$. 
Unfortunately, such a resonant enhancement would require 
$\gamma$-lasers, which do not exist at present.

Let us estimate the enhancement factor due to the relaxation of the 
forbiddeness of the $\beta$-transition in low-frequency fields. In the absence 
of the field, the matrix element of unique first-forbidden $\beta$-decay is 
$M_0\sim k_0 R/\hbar$, where, as usual, $k_0=\sqrt{2m \varepsilon_0}$. From 
eq.~(\ref{eq:Mind}) we then obtain 
\be
\frac{M_{ind}}{M_0}\,\simeq\, \frac{e E_0 \hbar}{\sqrt{2m \varepsilon_0}
\,\Delta \varepsilon_{1,2}}\,\equiv\, \chi_*\,,
\label{eq:factor}
\ee
i.e.\ the parameter that governs the enhancement of the probability due 
to the lifting of the forbiddeness is $\chi_*=\chi (2\varepsilon_0/
\Delta \varepsilon_{1,2})$. For the rate of the process in the case of 
relatively weak fields we then expect
\be
W\simeq W_0(\varepsilon_0)[1+a\chi^2 + b \chi_*^2 +\dots]\,,
\label{eq:W20} 
\ee
where $a$ and $b$ are numerical constants. 

Let us now sketch the calculation of the decay rate (the details can be found 
in ref.~\cite{Akh2}). 
Although all physical observables are gauge invariant, a judicious choice 
of the gauge can simplify the calculations greatly. For calculating the 
probability of forbidden $\beta$-decay in external electromagnetic fields, 
the scalar gauge (\ref{eq:4vpot1}) turns out that to be most convenient. 
The reason for this is that the corrections to the nuclear wave functions can 
then be described in the lowest order in the small parameter $e E_0 R/\Delta 
\varepsilon_{1,2}$ (see (\ref{eq:Mind})). At the same time, the corresponding 
interaction parameter in the Coulomb gauge, %
\footnote{Here we consider the parameter describing the modification of 
the wave function of the parent nucleus. The daughter nucleus can be 
considered similarly.} 
\be
\frac{e}{c}\frac{A_0\ p_{ni}}{M}
\frac{1}{ (\varepsilon_{n}-\varepsilon_{i})} =
\frac{e E_0 d_{ni}}{\hbar\omega}\,,
\label{eq:param}
\ee
is not in general small in low-frequency fields, and therefore it cannot be 
used as an expansion parameter in perturbation-theory calculations. 
(Here $A_0=cE_0/\omega$ is the amplitude of the vector-potential and 
$n$ corresponds to an excited nuclear state). Moreover, 
unlike the scalar-gauge parameter $e E_0 R/\Delta \varepsilon_{1,2}$, it is 
not suppressed for large values of $\varepsilon_{n}-\varepsilon_{i}$; 
this means that in calculating nuclear wave functions in the presence of 
external fields one cannot constrain oneself to the contributions of only 
lowest-lying excited states and has to sum over the whole spectrum of 
nuclear excitations. Therefore, we choose to calculate in the scalar gauge 
(\ref{eq:4vpot1}).

The wave function of a non-relativistic electron in the scalar gauge 
takes the form~\cite{keldysh}
\be
\Psi_{\vec{k}}'(\vec{r}, t)=\exp\left\{\frac{i}{\hbar}[\vec{k}-(e/c)\vec{A}(t)] 
\vec{r}-\frac{i}{2m\hbar}\int^t [\vec{k}-(e/c)\vec{A}(t')]^2\,d t' \right\}, 
\label{eq:Keld2} 
\ee
where the prime refers to the quantities in the scalar gauge. Note that  
here $\vec{A}(t)$ does not have the meaning of a vector-potential, which 
is zero in the scalar gauge (\ref{eq:4vpot1}); by definition, 
in eq.~(\ref{eq:Keld2}) $\vec{A}(t)\equiv -c\int^t\!dt' \vec{E}(t')$. 

Comparing the electron wave functions (\ref{eq:Keld1}) and (\ref{eq:Keld2}), 
we find that they are related by 
\be
\Psi_k'(\vec{r}, t)=e^{-\frac{i e}{\hbar c}\vec{A}(t)\vec{r}}
\Psi_k(\vec{r},t)\,.
\label{eq:Keld3}
\ee
This can be readily understood by noting that the gauge transformation 
from the Coulomb gauge (\ref{eq:4vpot}) to the scalar gauge (\ref{eq:4vpot1}) 
is $A^\mu (t,\vec{x}) \to A^\mu (t,\vec{x})+\partial^\mu 
\eta(t,\vec{x})$ with the gauge function $\eta(t,\vec{x})=\vec{A}(t)\vec{r}$. 
The wave functions of particles of charge $e$ should then transform as  
\be
\psi(t, \vec{r})\to 
e^{-(i e/\hbar c)\eta(t, \vec{r})}\psi(t, \vec{r})\,. 
\label{eq:gauge}
\ee
The electron wave functions in the Coulomb gauge and scalar gauge thus have 
different coordinate dependence. For allowed $\beta$-transitions the electron 
wave function inside the nucleus is replaced by a constant, and therefore it 
does not matter whether the Coulomb-gauge or scalar-gauge expression is used. 
For calculating the rate of forbidden $\beta$-decays the wave function 
(\ref{eq:Keld2}) should be employed.

The calculations \cite{Akh2} confirms the expected result (\ref{eq:W20}) for 
the rate of unique first-forbidden $\beta$ decay in the field of a strong 
electromagnetic wave. Since nuclear excitation energies $\Delta 
\varepsilon_{1,2}$ are typically of the same order of magnitude as the energy 
release $\varepsilon_0$, and in all known cases are not smaller than $\sim 1$ 
keV, for currently attainable fields the forbiddeness lifting term $b\chi_*^2$ 
in (\ref{eq:W20}) is extremely small, as is the phase space enhancement 
term $a\chi^2$. We discuss this point in more detail in the next section. 
Notice, however, that the numerical value of the coefficient $a$ in 
(\ref{eq:W20}) is significantly larger than in the case of allowed 
$\beta$-transitions: for unique first-forbidden $\beta$-decays the 
calculation \cite{Akh2} yields $a=315/8$, to be compared with 35/8 in 
eq.~(\ref{eq:W12}).

\section{\label{sec:disc}Summary and discussion}
We have considered, both qualitatively and quantitatively, the influence 
of the field of a strong electromagnetic wave on the characteristics of 
quantum processes with participation of non-relativistic charged 
particles. Our qualitative analysis has shown that the parameter that 
determines the influence of the external field on the differential 
characteristics of the processes (such as energy spectra and angular 
distributions of the final-state particles) is $\xi=e E_0/(\sqrt{2 m 
\varepsilon_0}\,\omega)$. The processes of modification of these 
differential characteristics in external fields are essentially 
of classical nature. 
	
At the same time, the parameter that governs the modification of the total 
probabilities of the processes (such as decay rates and total cross sections) 
is $\chi=e E_0 \hbar/(\sqrt{2 m \varepsilon_0} \,2 \varepsilon_0)$. It 
describes the energy $\delta\varepsilon_D$ obtained by the charged particle 
from the field (or given to the field) over the distances of order of the 
particle's de~Broglie wavelength. In relatively weak fields, when $\chi \ll 1$, 
this energy is given by $\delta\varepsilon_D \simeq \chi \varepsilon_0$, 
whereas in the strong field limit ($\chi\gg 1$) one has  $\delta
\tilde{\varepsilon}_D 
\simeq \chi^{2/3} \varepsilon_0$. The fractional power of $\chi$ enters 
in the latter case because of the field dependence of the charged particle's 
de~Broglie wavelength. The process of energy exchange between the produced 
particle and the external field in the process of the formation of the particle 
(and therefore the modification of the total probability of the process)     
is inherently quantum in its nature.

We also estimated the total probabilities of quantum processes in external 
electromagnetic fields, considering nuclear $\beta$-decay as an example. 
Our estimates were made in the limits $\chi \ll 1$ and $\chi \gg1$. In the 
latter case the decay rate for a $\beta$-active nucleus ($\varepsilon_0>0$) 
and the rate of the field-induced $\beta$-decay of the daughter nucleus which 
is stable in the absence of the external field ($\varepsilon_0<0$)
are practically the same. Examples of such transitions are 
$^3{\rm H} \to {}^3{\rm He} + e^- + \bar{\nu}_e$ ($\varepsilon_0>0$) and  
$^3{\rm He} \to {}^3{\rm H} + e^+ + \nu_e$ ($\varepsilon_0<0$). In the 
weak-field limit the induced decay in the case $\varepsilon_0<0$ is predicted 
to be exponentially suppressed in the low-frequency (tunneling) limit 
$\omega t_2 \ll 1$, where $t_2$ is the tunneling time defined in 
(\ref{eq:cond1}), and power-suppressed in the multi-photon regime 
$\omega t_2 \gg 1$.  

The dependence of the total probabilities of quantum processes on the field 
frequency $\omega$ comes through the expansion in powers of the parameter 
$(\omega t_x)^2$, where $t_x$ is the characteristic formation time of the 
produced charged particle. At the same time, as was mentioned above, the energy 
obtained by the particle in the process of its formation is given by the work 
of the field on the particle over the distance of order of the particle's 
formation length $l_x$: $\delta\varepsilon\simeq e E_0 l_x$. The quantities 
$l_x$ and $t_x$ are different in different regimes:    
\begin{itemize}
\item
Weak fields : 
\[
\varepsilon_0~>~0:
~~\quad l_x\simeq 
\frac{\hbar}{2\sqrt{2m\varepsilon_0}}\,, \qquad 
t_x\simeq\hbar/2\varepsilon_0 \qquad~~~~~{\Rightarrow} 
\quad~~~ \omega t_x~=~\hbar\omega/2\varepsilon_0
~=~1/\delta\,,~~\qquad~~~
\]
\[
\varepsilon_0~<~0: \quad ~~l_x\simeq 
\frac{|\varepsilon_0|}{e E_0}\,, \qquad
\quad~~~\,t_x=
\frac{\sqrt{2m |\varepsilon_0|}}{e E_0}  \quad~~~
{\Rightarrow} \quad~~~ \omega t_x~=~\sqrt{2m 
|\varepsilon_0|}\omega/e E_0 
~=~1/\xi\,,\;
\]

\item
Strong fields :
\[
\qquad\quad l_x\simeq \left(\frac{\hbar^2}{8m e E_0}\right)^{1/3},  
~~~\quad~~
t_x\simeq \left(\frac{m\hbar}{e^2 E_0^2}\right)^{1/3}
\quad\!
{\Rightarrow} \quad ~~\omega t_x~=~(\hbar\omega/ 
2\varepsilon_0)\chi^{-2/3}\,.
\qquad\qquad\qquad
\]

\end{itemize}

We have also discussed a simple method of calculating the total probabilities 
of quantum processes with participation of non-relativistic charged 
particles in the field of an electromagnetic wave, considering nuclear 
$\beta$-decay as an example.   
The method does not rely on a summation of partial probabilities with 
absorption from the field or emission into it of all possible 
numbers of photons; instead, the total probabilities are calculated 
directly. The results of the direct calculations in the cases of allowed and 
forbidden $\beta$-decay fully confirmed the estimates made in  
sec.~\ref{sec:qualit}, often even including the values of numerical 
coefficients. 

How large can actually the effects of strong electromagnetic fields on 
$\beta$-decay rates be? The parameter $\chi$ that determines the modification 
of the rates can be written as 
\be
\chi=\frac{e E_0 \hbar}{\sqrt{2 m \varepsilon_0} \,2 \varepsilon_0} 
=\frac{E_0}{E_c}\left(\frac{m c^2}{2\varepsilon_0}\right)^{3/2},
\label{eq:crit1}
\ee
where 
\be
E_c=\frac{m^2 c^3}{e\hbar}=1.323\times 10^{16}~\,{\rm V/cm}
\label{eq:crit2}
\ee
is the QED critical field strength (the so-called Schwinger field). The 
most powerful present-day lasers can reach the field intensities up to 
$I\sim 10^{22}~{\rm W/cm}^2$, and in the near future probably the 
intensities $I\sim 10^{24}~{\rm W/cm}^2$ will become available. From 
the formula  
\be
E_0({\rm V/cm})~\simeq~ 20\,\sqrt{I({\rm W/cm^2})} 
\ee
we then find that this corresponds to the field strengths $E_0/E_c\sim 
10^{-4}$ -- $10^{-3}$. For tritium $\beta$-decay from 
eq.~(\ref{eq:crit1}) we then obtain 
\be
\chi\simeq 5\times 10^{-3} ~-~5\times 10^{-2}\,.
\label{eq:range} 
\ee
Taking into account that for small $\chi$ the correction to the decay 
rate is of order $\chi^2$, we see that the field effect on tritium 
$\beta$-decay is too small to be observable. 

Since $\chi^2$ scales as $\varepsilon_0^{-3}$, one can expect significantly 
stronger effects for $\beta$-decays with smaller energy release. 
In particular, for unique first forbidden $\beta$-decay 
$^{187}{\rm Re}(\frac{5}{2}^+)\to {}^{187}{\rm Os}(\frac{1}{2}^-)$ one 
has  $\varepsilon_0\simeq 2.64~{\rm keV}$, and the parameter $\chi$ can 
formally take values $\chi\sim 1$ even for present-day lasers. Does that 
mean that we can already observe strong effects of the laser fields on 
nuclear $\beta$-decay?

Unfortunately, in reality the situation is not that promising at the 
moment. First, for nuclei with small energy release (and especially for 
forbidden $\beta$-decays) the decay rates are extremely small, and even 
sizeable corrections to them are not easily observable: the lifetime of 
$^{187}{\rm Re}$, for example, is $\sim 5\times 10^{10}$ yr. The 
situation is complicated by the fact that the powerful lasers have very low 
pulse duration (in the femtosecond range) and repetition frequency 
which is typically $\sim 10^{-3}~{\rm Hz}$. Finally, except for 
$\omega \gg \omega_{at}$ where $\omega_{at}$ are characteristic atomic 
frequencies, the atomic electrons would screen the external fields, 
greatly reducing their strengths at the nucleus, and to observe the 
field effect one would first have to ionize the atom of the $\beta$-active 
element.  

In contrast to this, laser field effects on energy spectra and angular  
distributions of electrons and positrons emitted in nuclear $\beta$-decay can 
be quite significant.  
In addition, strong electromagnetic fields can still influence sizeably 
the total probabilities of atomic and molecular processes with emission 
of low-energy electrons. An example of such a process is a photo-ionization  
in two fields, one of which is a weak field with the energy of 
the quantum $\hbar \Omega$ slightly above (or slightly below) the ionization 
potential $I$ and the other -- an intense field with the frequency 
satisfying $\hbar\omega \ll I$.

\vspace*{2mm}
{\em Acknowledgments}. At different times, the author benefited from very  
useful discussions with W.~Becker, A.M.~Dykhne, Yu.V. Gaponov, 
V.A.~Khodel and M.E.~Shaposhnikov. This work was supported in part by 
the Sonderforschungsbereich TR 27 of the Deutsche Forschungsgemeinschaft.

\appendix
\renewcommand{\theequation}{\thesection\arabic{equation}}
\appsection
\renewcommand{\thesection}{\Alph{section}}
\section*{Appendix \Alph{section}:
Validity of approximations}
We consider here the validity conditions for the approximations adopted in 
our calculations of the total probabilities of quantum processes. 

The dipole approximation implies that the wavelength of the external 
electromagnetic field is assumed to be large compared to the characteristic 
length parameters $l_x$ involved in the problem: $\lambda~=~2\pi c/\omega~
\gg~l_x$. Let us consider this condition in various regimes.

\vspace*{0.2cm}
(a) ~{$\chi\ll 1$, ~$\varepsilon_0>0$}: 
~~$l_x\simeq \lambdabar_D/2=\hbar/(2\sqrt{2m \varepsilon_0})$. The dipole 
approximation requires 
\be
\frac{\hbar\omega}{2\varepsilon_0} ~\ll~ 4\pi \left(\frac{m c^2}
{2\varepsilon_0}\right)^{1/2}\gg 1\,,
\label{eq:cond2}
\ee
which is certainly satisfied for $\hbar\omega \lesssim \varepsilon_0$.  

(b) ~{$\chi\ll 1$, $\varepsilon_0<0$}: ~$l_x~=~|\varepsilon_0|/e E_0$. 
The dipole approximation is valid provided that   
\be
\frac{\hbar\omega}{2|\varepsilon_0|}~\ll~4\pi \chi 
\left(\frac{m c^2}{2|\varepsilon_0|}\right)^{1/2}.
\label{eq:cond3}
\ee

(c) ~{$\chi\gg 1$, $\varepsilon_0>0$} ~or ~$ <0$: 
~$l_x~=~[\hbar^2/(8m e E_0)]^{1/3}$. The validity condition is 
\be
\frac{\hbar\omega}{2|\varepsilon_0|}~\ll~4\pi \chi^{1/3} 
\left(\frac{m c^2}{2|\varepsilon_0|}\right)^{1/2},
\label{eq:cond4}
\ee
which is satisfied with a large margin.

Next, we discuss the non-relativistic approximation. In the weak-field limit 
($\chi\ll 1$) the standard condition is  
\be
\varepsilon_0~\ll~m c^2. 
\label{eq:cond5}
\ee
However, in the strong field limit $\chi\gg 1$ one also has to make sure that 
the energy obtained from the field by the charged particle during its 
formation does not take it out of the non-relativistic domain: 
$ e E_0 l_x~\ll~m c^2$. Taking into account that in the strong-field regime 
$l_x~=~[\hbar^2/(8m e E_0)]^{1/3}$, we arrive at the condition  
\be
E_0~\ll~E_c~=~1.323\times 10^{16}~{\rm V/cm}. 
\label{eq:cond6}
\ee

In our calculations of the rates of nuclear $\beta$-decay in external fields 
we were neglecting the final-state interaction of the produced electron or 
positron with the Coulomb field of the nucleus. This is permissible when the 
Coulomb energy $Ze^2/l_x=\alpha Z \hbar c/l_x$ is small compared to the 
characteristic energy $\varepsilon_x$ of the process. Consider now the 
Coulomb parameter $\alpha Z\hbar c/l_x \varepsilon_x$ in various regimes. 

(a) ~{$\chi\ll 1$, ~$\varepsilon_0>0$}: 
~~$l_x\simeq \hbar/(2\sqrt{2m \varepsilon_0})$, 
~~$\varepsilon_x\simeq\varepsilon_0$. ~In this case    
\be
\alpha Z\hbar c/\varepsilon_x l_x~\simeq~4\alpha Z c/v_0~=~
4\alpha Z c(m/2\varepsilon_0)^{1/2}\,,
\label{eq:cond7}
\ee
which is (up to the factor 4) just the standard Coulomb parameter. As an 
example, for $\beta$-decay of $^3$H we have $Z=2$, $(m/2\varepsilon_0)^{1/2}
~\simeq ~3.7$, $\alpha Z \hbar c/\varepsilon_x l_x~\simeq~0.2$.  

(b) ~$\chi\ll 1$, $\varepsilon_0<0$: 
~$l_x~=~|\varepsilon_0|/e E$, ~$\varepsilon_x=|\varepsilon_0|$, 
\be
\alpha Z\hbar c/\varepsilon_x l_x~\simeq~4 \alpha Z c 
(m/2|\varepsilon_0|)^{1/2}\,\chi\,. 
\label{eq:cond8}
\ee

(c) ~$\chi\gg 1$, $\varepsilon_0>0$} ~or ~$ <0$: 
~$l_x~=~[\hbar^2/(8m e E_0)]^{1/3}$, 
~~$\varepsilon_x=\chi^{2/3}|\varepsilon_0|$\,, 
\be
\alpha Z\hbar c/\varepsilon_x l_x~\simeq~4 \alpha Z(m/2|\varepsilon_0|)^{1/2}
\,\chi^{-1/3}\,.
\label{eq:cond9}
\ee

\renewcommand{\theequation}{\thesection\arabic{equation}}
\appsection
\renewcommand{\thesection}{\Alph{section}}
\section*{Appendix \Alph{section}:
Some useful formulas}
Here we collect some formulas which have been used in calculation done 
in secs.~\ref{sec:allowed} and \ref{sec:forbid}. 

An integral representation for $J_0(z)$:
\be
J_0(z)~=~\frac{1}{\pi}\int_0^\pi e^{i z \cos\theta}\,d\theta\,.
\ee

\noindent
Gradshteyn \& Ryzhik \cite{GrRyzh}, 6.631(6):
\be
\int_0^\infty x^{\nu+1}e^{\pm i \alpha x^2} J_\nu(\beta x)
\,dx~=~\frac{\beta^\nu}{2\alpha^{\nu+1}} \exp\left[\pm i \left(\frac{\nu+1}{2}
\pi-\frac{\beta^2}{4\alpha}\right)\right].
\ee
$[\alpha>0,~-1<{\rm Re}\nu < 1/2,~\beta>0]$.

\vspace*{4mm}
\noindent
Gradshteyn \& Ryzhik \cite{GrRyzh}, 3.382.7:
\be
\int_{-\infty}^\infty \frac{e^{-i p x}}{(\beta-i x)^{\nu}}\,dx
~=~\left\{\begin{array}{c}
\frac{2\pi p^{\nu-1} e^{-\beta p}}{\Gamma(\nu)},~~p>0 \\
~~~~~~0,\qquad\quad p<0 
\end{array}\right.\,.
\ee
$[{\rm Re}\nu >0,~{\rm Re}\beta >0]$.

\noindent
For an electromagnetic wave with the electric field strength  
$\vec{E}(t)=\{E_0 \sin\omega t, ~{}-E_0 \cos\omega t, ~~0\,\}$ the 
electron  
wave function in the Coulomb gauge~(\ref{eq:Keld1}) is  
\be
\Psi_{\vec{k}}(\vec{r},\,t)=\exp\left\{\frac{i}{\hbar}\vec{k}\vec{r}
-\frac{i}{\hbar}\left[\left(
\frac{k^2}{2m}+\frac{e^2 A_0^2}{2m c^2}\right)t-\frac{e A_0}{m c \omega}
(k_x\sin\omega t-k_y\cos\omega t)\right]\right\},
\label{eq:Keld4}
\ee
where $A_0=E_0 c/\omega$.


\end{document}